\documentclass[preprint2]{aastex62}

\graphicspath{{./}}

\begin{document}

\title{A High Spectral Resolution Study of the Soft X-ray Background with the X-ray Quantum Calorimeter}

\correspondingauthor{Dallas Wulf}
\email{dallas.wulf@mcgill.ca}
\author{Dallas Wulf}
\altaffiliation{Current Address: McGill University Montreal, QC\\H3A 2T8, Canada}
\affil{University of Wisconsin at Madison \\
Madison, WI 53706, USA}

\author{Megan E Eckart}
\altaffiliation{Current Address: Lawrence Livermore National\\Laboratory Livermore, CA 94550, USA}
\affil{NASA/Goddard Space Flight Center \\
Greenbelt, MD 20771, USA}

\author{Massimiliano Galeazzi}
\affil{University of Miami \\
Coral Gables, FL 33146, USA}

\author{Felix Jaeckel}
\affil{University of Wisconsin at Madison \\
Madison, WI 53706, USA}

\author{Richard L Kelley}
\affil{NASA/Goddard Space Flight Center \\
Greenbelt, MD 20771, USA}

\author{Caroline A Kilbourne}
\affil{NASA/Goddard Space Flight Center \\
Greenbelt, MD 20771, USA}

\author{Kelsey M Morgan}
\altaffiliation{Current Address: National Institute of Standards and\\Technology, Boulder, CO 80305, USA}
\affil{University of Wisconsin at Madison \\
Madison, WI 53706, USA}

\author{Dan McCammon}
\affil{University of Wisconsin at Madison \\
Madison, WI 53706, USA}

\author{F Scott Porter}
\affil{NASA/Goddard Space Flight Center \\
Greenbelt, MD 20771, USA}

\author{Andrew E Szymkowiak}
\affil{Yale University \\
New Haven, CT 06520, USA}



\begin{abstract}\label{sec:abst}

We present here a combined analysis of four high spectral resolution observations of the Diffuse X-ray Background (DXRB), made using the University of Wisconsin-Madison/Goddard Space Flight Center X-ray Quantum Calorimeter (XQC) sounding rocket payload. The observed spectra support the existence of a $\sim$0.1~keV Local Hot Bubble and a $\sim$0.2~keV Hot Halo, with discrepancies between repeated observations compatible with expected contributions of time-variable emission from Solar Wind Charge Exchange (SWCX). An additional component of $\sim$0.9~keV emission observed only at low galactic latitudes can be consistently explained by unresolved dM stars. 

\end{abstract}



\section{Introduction}\label{sec:intro}
Observations of diffuse X-rays below 1~keV revealed the presence of widely-distributed hot gas in the Galaxy. We have come to realize that understanding the roles of this material in metal transport, stellar feedback, the disk-halo connection, and the exchange of matter with circumgalactic space is critical to central questions on the structure and evolution of galaxies. But useful observations for this purpose are difficult. 

The Diffuse X-ray Background (DXRB) below 1~keV originates from several sources ranging from solar to extragalactic distance scales.  With the exception of the Cosmic X-ray Background (CXRB), which can be represented by a power law, most of the DXRB flux in this energy range is in lines from highly charged ions.  Hot interstellar gas at temperatures ranging from $1\text{--}4\times10^{6}$~K produces thermal emission throughout the Galaxy and Galactic halo. Closer to home, highly charged ions originating from the solar wind emit X-rays through an electron capture process known as charge exchange, or in this case specifically, Solar Wind Charge eXchange (SWCX) \citep{cravens_comet_1997,koutroumpa_solar_2009,galeazzi_origin_2014}.  These two very different emission mechanisms will produce the same lines, making them difficult to distinguish observationally. However, observations capable of resolving closely spaced lines in this part of the spectrum can utilize the relative intensity ratios of two or more lines from the same ion to constrain the relative contribution from each to the DXRB (see details in Section~\ref{sec:cx}). 

To date, very few observations of the DXRB have had the spectral resolution required to resolve
individual emission lines. All of the maps with broad sky coverage were made with gas proportional counters, which have resolving powers, $R \equiv E/ \Delta E$, of $\sim$1 at 250~eV and $\sim$2.5 at 1~keV.  Slit-less grating spectrometers such as those on Chandra and XMM-Newton cannot be used on a very extended source such as the DXRB. Neither would adding slits help, since the surface brightness is too low to get an appreciable flux through a slit narrow enough to allow useful spectral resolution.  Therefore, spectra must be obtained with energy-resolving detectors such as CCDs or microcalorimeters.  Unfortunately, the statistics of charge formation fundamentally limit the resolving power of CCDs to $\sim$15 at 600~eV.  This is adequate to identify a few of the brightest lines between 500 and 1000~eV (e.g. O{\scriptsize ~VII}, O{\scriptsize ~VIII}, and Ne{\scriptsize ~IX}), but cannot resolve any of the more closely-spaced lines below $\sim$250~eV.  Microcalorimeters get around the charge statistics limit by converting all of the photon energy to heat and measuring the temperature rise of the detector produced by a single photon.

Here we present a combined analysis of four high spectral resolution observations of the DXRB made by the University of Wisconsin-Madison/Goddard Space Flight Center sounding rocket payload, the X-ray Quantum Calorimeter (XQC). XQC is a microcalorimeter X-ray spectrometer optimized for observing faint, extended sources like the DXRB below 1~keV with sufficient spectral resolution to identify individual emission lines. The paper proceeds as follows: Details about the sounding rocket observations and hardware configurations are given in Section~\ref{sec:obs}.  In Sections~\ref{sec:datRed} and \ref{sec:spec}, we describe the data reduction process and spectral analyses.  A discussion of the inferred properties of the DXRB can be found in Section~\ref{sec:discussion} and our conclusions in Section~\ref{sec:conclusion}.

\section{Sounding Rocket Observations}\label{sec:obs}
\begin{figure*}
\gridline{\fig{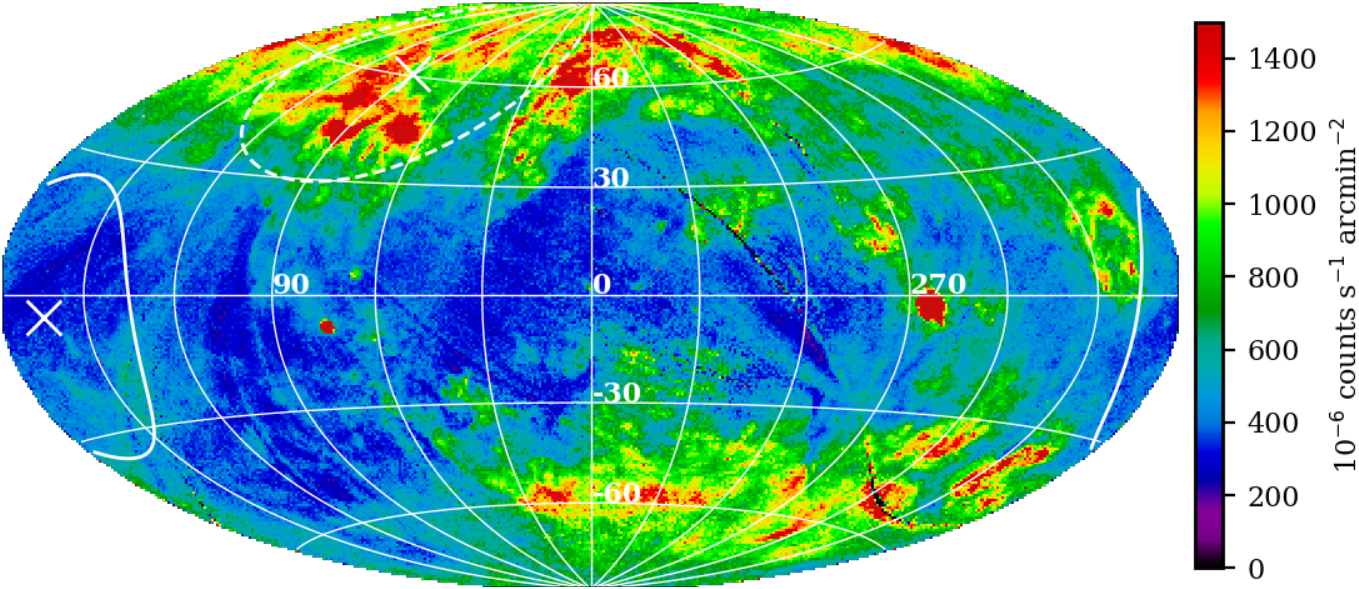}{0.63\textwidth}{1/4 keV (R12)}}\label{fig:14kevTarget}
\gridline{\fig{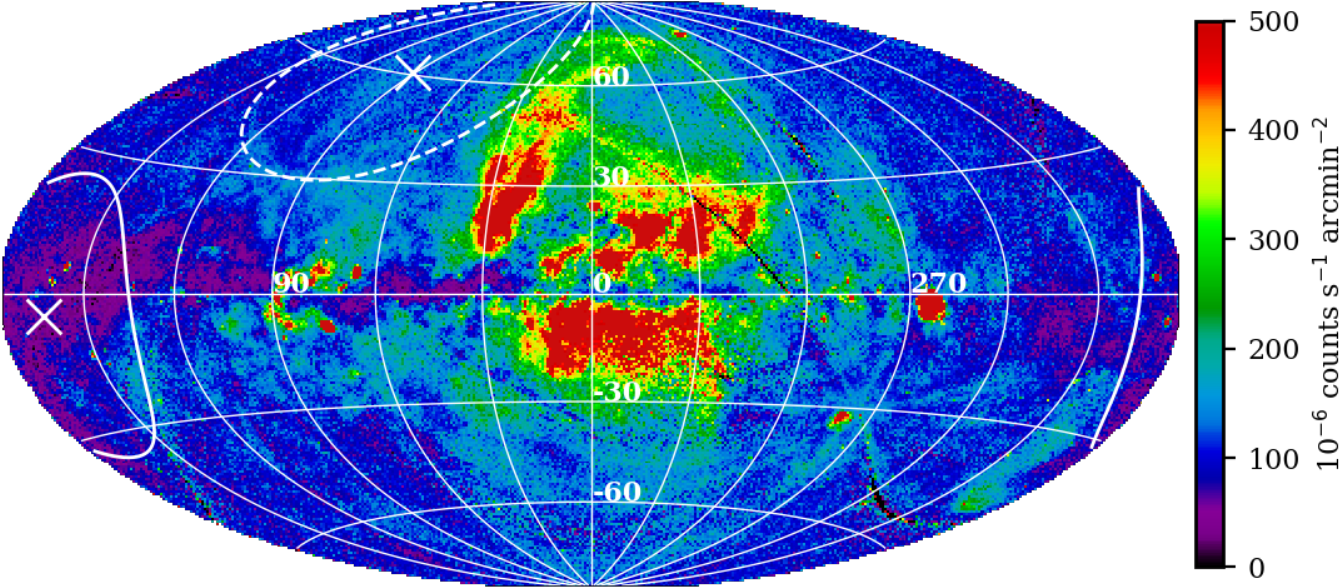}{0.63\textwidth}{3/4 keV (R34)}}\label{fig:34kevTarget}
\gridline{\fig{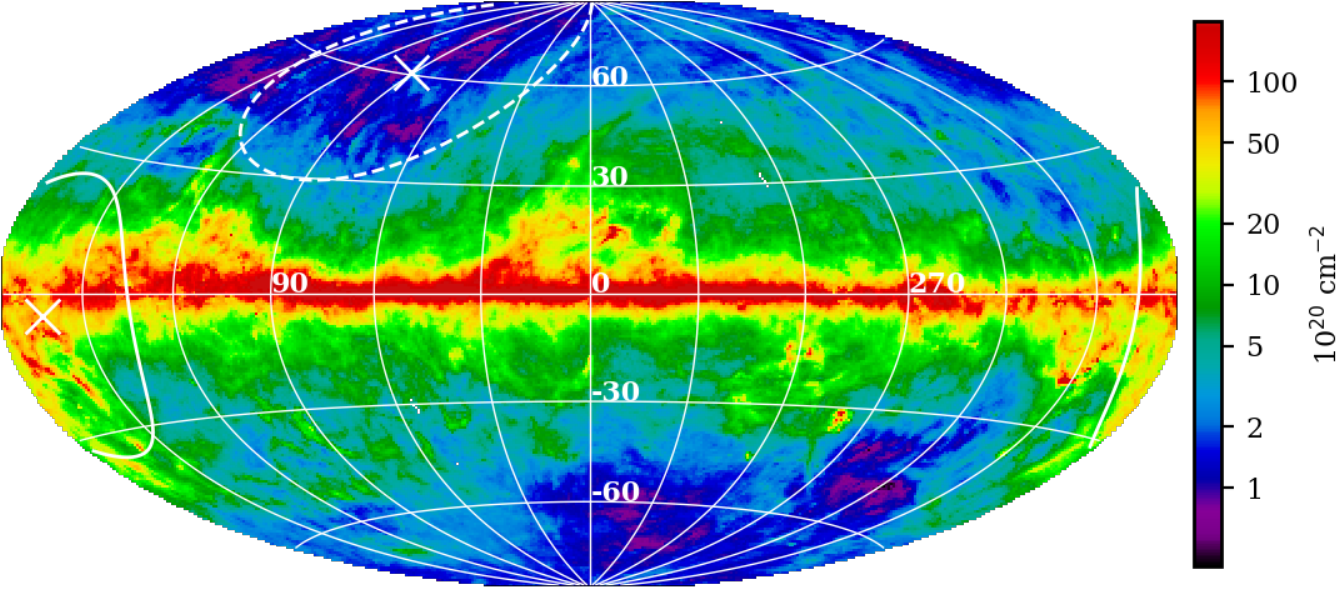}{0.63\textwidth}{IRAS-derived N$_{H}$}}\label{fig:irnhTarget}
\caption{\label{fig:targetMaps} Target area of XQC observations projected on all-sky maps. Maps are in Galactic coordinates with the plane of the Galaxy located along the equator and the center of the Galaxy at the center of the map. The $\sim$1~sr XQC fields of view are centered at $\ell,b=90\degr,+60\degr$ (dashed line) and $\ell,b=165\degr,-5\degr$ (solid line). ROSAT maps of R12 and R34 are from \cite{snowden_rosat_1997}, and IRAS map is from data in \cite{schlegel_maps_1998} converted to equivalent H{\scriptsize ~I} column.}
\end{figure*}

Table~\ref{ta:missions} summarizes the observing details of four XQC missions launched between 1999 and 2013.  All missions were launched from White Sands Missile Range (WSMR) in New Mexico, USA (lat, long = 32:40:00N, 106:20:00W). The first two missions, 27.041 and 36.223, observed at high Galactic latitude ($\ell,b=90\degr,+60\degr$), chosen to be typical of high latitudes while avoiding the bright enhancements of Loop I and the North Polar Spur.  The second two missions, 36.264 and 36.294, observed at low Galactic latitude ($\ell,b=165\degr,-5\degr$), avoiding emission associated with the Galactic center.  Projections of the target areas onto all-sky maps are depicted in Figure~\ref{fig:targetMaps}.  A previous analysis of the data from 27.041 was presented in \cite{mccammon_high_2002}, and data from 36.223 have been presented in \cite{crowder_observed_2012}.  Data from 36.264 and 36.294 are presented here for the first time.

\begin{table*} 
\caption{Summary of XQC sounding rocket observations.}
\label{ta:missions}
	\begin{center}
	{\footnotesize
	\begin{tabular}{c c c c c}
	\hline
	Mission & Time & & Exposure & Target\\
	Number & (UT) & Date & (s) & ($\ell,b$)\\
	\hline
	27.041 & 9:00 & 1999 Mar 28 & 100.7 & 90\degr,+60\degr\\
	36.223 & 5:30 & 2008 May 1& 176.9 & 90\degr,+60\degr\\
	36.264 & 7:00 & 2011 Nov 5 & 205.4 & 165\degr,--5\degr\\
	36.294 & 9:25 & 2013 Nov 3 & 150.0 & 165\degr,--5\degr\\
	\hline
	\end{tabular}
	}
	\end{center}
\end{table*}

Mission 27.041 featured the first generation detector format described in \cite{mccammon_high_2002}, while the following three missions featured the larger detector format described in \cite{mccammon_x-ray_2008}.  Compared to the first detector, the new detector provides a factor of four improvement in collecting area (1.44~cm$^{2}$ compared to 0.36~cm$^{2}$), as well as improved energy resolution ($\sim$6~eV~FWHM below 1~keV compared to $\sim$9~eV~FWHM). A hard landing destroyed the detector from Mission~36.223, but it was replaced by a similar detector for Missions~36.264 and 36.294. All four observations feature a $\sim$1~sr non-imaging field of view, providing large throughput motivated by the low surface brightness of the DXRB. A minimum of five thin  aluminized polyimide filters ranging in temperature from 0.05--130K were used to shield the detector plane from 300~K infrared radiation ($\lambda \gtrsim 2 \mu \text{m}$) while still having acceptable transmission to soft X-rays.  On the most recent flight, an outer sixth filter held at 300K was added to prevent contamination of the inner cold filters.  The issue of filter contamination is discussed more detail in Section~\ref{sec:contaminate}. For a more detailed description of the payload, detector, and infrared-blocking filters, the reader is referred to \cite{mccammon_high_2002,mccammon_x-ray_2008}.   

\section{Data Reduction}\label{sec:datRed}
Due to limited telemetry bandwidth, only triggered data segments containing X-ray events are telemetered to the ground station during an observation. Although almost 100\% of X-ray events are contained within this data set, segments containing more than one event cannot be processed with optimum spectral resolution, effectively reducing detector livetime. Therefore, beginning with Mission 36.223, on-board flash storage was installed, enabling retrieval of complete time series data for all pixels upon payload recovery. The complete time series data permits the use of an overlapped pulse-fitting algorithm \citep{wulf_technique_2016, wulf_technique_2019} which is capable of processing these closely spaced X-ray events without loss of resolution. Missions 36.223, 36.264, and 36.394 were each reprocessed with the most recent version of this algorithm.  No changes were made to the data reduction of 27.041 from \cite{mccammon_high_2002}.

\subsection{Gain and Linearity Corrections}\label{sec:gain}
The gain of a microcalorimeter changes with temperature, partly because the heat capacity of the absorber increases with temperature, but primarily because the temperature response of the thermometers is non-linear.  While the coldplate temperature can generally be precisely regulated at 50~mK, changes in the thermal environment can affect the temperature offset between the pixels and the coldplate, and thus impact the gain. To track these gain changes, a $^{41}$Ca calibration source continuously illuminates the entire array with K K$\alpha$ and K$\beta$ X-rays (3.31 and 3.59~keV, respectively).

The same non-linear thermometer response which is responsible for temperature dependent gain is also responsible for a non-linear energy response (i.e. as the temperature of the pixel changes during an X-ray event, the gain also changes).  In order to convert measured pulse height to energy, a cubic polynomial is fit to calibration data containing emission lines of known energy.  For this purpose, there is a multi-target fluorescent calibration source installed on the inside of the gate-valve slide which emits a range of characteristic X-ray lines between 183 and 1740~eV, and which is visible to the detector whenever the gave-valve is closed. For each mission, calibration data for establishing the non-linear energy scale were collected during $\sim$30 minutes immediately preceding the launch.

\subsection{Event Selection}\label{sec:events}
The overlapped pulse fitting algorithm is capable of processing nearly 100\% of X-ray events.  However, high energy events that cause ADC saturation ($\gtrsim$10~keV) cannot be analyzed, and are therefore excised from the data streams prior to processing.  Any other X-ray events that are within the optimal filter length of a saturating event must also be removed. This must be done to ensure that the remaining events are all processed with optimum resolution.  This cut dominates the processing deadtime for the three observations which are processed this way.  Additionally, an event is excluded from the final analysis if it arrives within 10~ms following another event on the same pixel, and two events are excluded if they arrive within 2.5~ms of each other on adjacent pixels.  The first condition removes events which are subject to rapidly changing gain due to the pixel cooling from the previous event. The second condition removes cosmic ray events and other sources of thermal cross-talk that deposit energy in neighboring pixels, resulting in events that are otherwise indistinguishable from true X-ray events. These event selection intervals can be compared with the 1.5~ms and 9~ms typical rise and fall times of the detectors, and they introduce negligible additional deadtime.

Finally, all the events from a single pixel are excluded if the arrival time distribution of events on that pixel is significantly different than a random distribution described by a single rate, or if the energy distribution (i.e. the spectrum) is significantly different than that of the entire detector array.  The spectral comparison is made using the $K$-Sample Anderson-Darling test for equality \citep{scholz_k-sample_1987}, chosen because it is non-parametric (i.e. requires no assumptions about the underlying data distribution), and has sensitivity at the tails of the distribution.  Pixels were considered to fail this test if the null hypothesis was rejected at the 95\% level.  These tests are an added safeguard to ensure that time- and pixel-dependent gain corrections do not introduce any distortions to the final spectrum.  Depending on the observation, these tests affected between 3 and 5 pixels.

\subsection{Filter Contamination}\label{sec:contaminate}
Beginning with Mission 36.223, water ice contamination to the infrared-blocking filters became an issue.  Calibration data from immediately before and after the observation confirmed that the contamination formed during the time that the gate-valve was open, suggesting that the source of the contamination was outside the cryostat. Between Missions 27.041 and 36.223, a hermetic bulkhead was removed between the experiment section and the other sections of the payload.  The source of the contamination was initially assumed to be the payload venting through the experiment section. For Mission 36.264 therefore, the payload was purged with dry nitrogen gas prior to launch.  Unfortunately, this remedy failed to prevent water ice from accumulating on the filters, with total ice accumulation actually being worse than the previous mission.  For Mission 36.294, a sixth outermost filter held at 300K was added.  While the additional filter reduced throughput slightly, it successfully prevented ice from forming on the filters, resulting in a net improvement to throughput.  The exact source of water vapor responsible for the ice is still unknown.

For Missions 36.223 and 36.264, the correction for contaminated filter transmission is based on a combination of post-flight data (taken on the parachute after re-entry into the atmosphere) and in-flight data.  The same gate-valve calibration source described in Section~\ref{sec:gain} used to establish the non-linear gain scale prior to flight also provides a relative measurement of filter transmission following the observation.  These data have the advantage of measuring the transmission at well-defined energies, but sample a relatively small fraction of the filter area and only provide an estimate of the total ice accumulated at the end of the observation.  To complement this measurement, the on-sky spectrum measured throughout the observation is compared to predictions made by the ROSAT All-Sky Survey (RASS) data. In contrast to the gate-valve data, these data sample the entire filter area and provide time-domain information about ice accumulation, though measure only the RASS band-averaged transmission. For both flights affected by ice contamination, self-consistent models for filter transmission are selected using both of these measurements.

\subsection{Non X-ray Background}\label{sec:background}
The primary source of background events in the 100--1100~eV range is higher energy X-rays that have lost energy due to photoelectric escape.  Whenever an X-ray is stopped near the surface of the pixel, there is some probability that it will eject an electron from the surface via the photoelectric effect.  This happens $\sim$5\% of the time.  The energy carried away by the electron is lost, resulting in a lower observed energy.  The resulting spectrum can be approximated as flat between zero and the incident X-ray energy.  For all four XQC observations, the internal $^{41}$Ca calibration source at 3.3 and 3.7~keV is responsible for most of this background, contributing 0.5~counts~s$^{-1}$~keV$^{-1}$ to the 36.223 spectrum and 0.3~counts~s$^{-1}$~keV$^{-1}$ to the other 3 spectra (the calibration source intensity was greater for 36.223).

We also consider possible particle sources of backgrounds.  For primary cosmic ray protons, the minimum ionizing energy loss in a pixel is  $\gtrsim$5.5~keV, and thus does not contribute to the low energy spectrum directly.  However, cosmic rays can also eject electrons from surfaces surrounding the detector array which can produce events at lower energies.  We have looked for evidence of background from such cosmic ray secondaries when the gate-valve is closed, but the observed spectrum has always been consistent with the spectrum taken with the gate-valve closed at sea level, suggesting that this background is negligible.  

Another possible source of low-energy particle background is auroral electrons.  This has always been a concern for XQC, whose wide field of view prevents the use of magnetic brooms to keep electrons from entering the detector. Fortunately, with the exception of Mission 36.294, the data do not support a significant contribution from an electron background. This determination is made based on comparisons with the ROSAT survey data, which provide an approximate shape of the DXRB spectrum.  Only for Mission 36.294 do we observe an excess flux of approximately 0.4~counts~s$^{-1}$~keV$^{-1}$ in each of the ROSAT energy bands ($\sim$100--2500~eV). Auroral electrons provide a natural explanation for such a flat excess, as the energy loss trough the infrared-blocking filters ($\sim$3~keV, depending on energy) serves to flatten almost any initial energy distribution.

\subsection{Cross Check}\label{sec:check}
As a consistency check on the absolute flux, the final event selection is compared to the band fluxes measured by the RASS.  The energy resolution of XQC allows us to make a model-independent comparison by simply multiplying the observed spectra by the ratio of the ROSAT PSPC response to the XQC response at the energy observed by XQC.  The relative fluxes and band energies are listed in Table~\ref{ta:check}.  The reader is referred to \cite{snowden_rosat_1997} for a complete description of the ROSAT PSPC broad energy bands. For the three most recent missions, we find good agreement with the ROSAT data, especially at higher energies. Notably, we find the largest discrepancy in the R2 band for all observations. As will be discussed later in Section~\ref{sec:discussion}, this discrepancy can be explained by variability in the contribution from heliospheric SWCX. For Mission 36.264, there is also a discrepancy observed in the R4 and R5 bands, though this is also the mission with the most significant filter contamination from water ice.  As bands R4 and R5 are situated above the oxygen K-edge, they are especially sensitive to the temporal and spatial distributions of ice accumulation which may not be captured by the model for ice transmission. Relative fluxes for Mission 27.041 have been previously reported in \cite{mccammon_high_2002}

\begin{table*} 
\caption{Observed flux for all four missions relative to the RASS energy bands.  R1 is not included due to large statistical uncertainties.  Statistical uncertainties are given at the 1$\sigma$ level.}
\label{ta:check}
	\begin{center}
	{\footnotesize
	\begin{tabular}{l c c c c c c}
	\hline
	& \multicolumn{6}{c}{Relative ROSAT Band Flux (\%)} \\
	Mission & R1 & R2 & R4 & R5 & R6 & R7 \\
	\hline
	27.041$^{\rm a}$ & - & 143$\pm$12 & 94$\pm$7 & 81$\pm$6 & 69$\pm$6 & 89$\pm$11 \\
	36.223 & - & 85$\pm$14 & 97$\pm$7 & 91$\pm$5 & 99$\pm$5 & 107$\pm$7 \\
	36.264 & - & 65$\pm$16 & 76$\pm$10 & 81$\pm$7 & 99$\pm$7 & 106$\pm$7 \\
	36.294 & - & 75$\pm$9 & 92$\pm$7 & 92$\pm$6 & 109$\pm$8 & 117$\pm$10 \\
	\hline
	Energy$^{\rm b}$ (eV) & 110--284 & 140--284 & 440--1010 & 560--1210 & 730--1560 & 1050--2040 \\
	\hline
	\multicolumn{7}{l}{$^{\rm a}$ Values from \cite{mccammon_high_2002}.} \\
	\multicolumn{7}{l}{$^{\rm b}$ 10\% of peak response \citep{snowden_rosat_1997}.} \\
	\end{tabular}
	}
	\end{center}
\end{table*}

\section{Spectral Analysis}\label{sec:spec}
Spectral fitting was performed with XSPEC version 12.9.0, using the PyXspec interface \citep{arnaud_xspec:_1996}.  Thermal emission was modeled with APEC version 2.0.2 \citep{smith_collisional_2001} using solar abundances from \cite{anders_abundances_1989}.  Depleted abundances from \cite{savage_interslar_1996} were also considered, but were not found to improve the fits. Charge exchange emission was modeled with AtomDB's Charge eXchange model, VACX, version 1.0.1 \citep{smith_approximating_2012}.  Following \cite{smith_resolving_2014}, two SWCX components representing contributions from fast and slow solar wind were used, with relative abundances taken from Table~1 in \cite{von_steiger_composition_2000}.  

\begin{figure}
\plotone{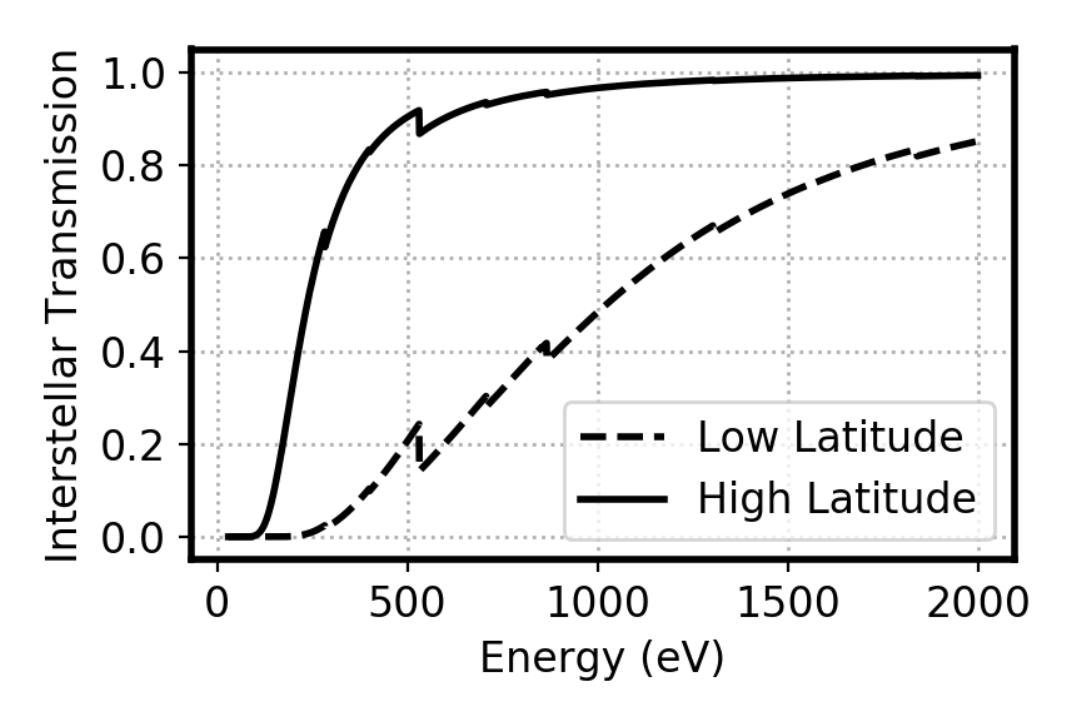}
\caption{\label{fig:phabs} Interstellar transmission averaged over XQC fields of view.  This transmission is used to account for interstellar absorption of the Cosmic X-ray Background (CXRB) and hot Galactic halo emission in spectral models.}
\end{figure}

Interstellar absorption of the Cosmic X-ray Background (CXRB) and hot Galactic halo was calculated from the transmission-weighted average over the full XQC field of view.  The column depth of the Galaxy in each direction was derived from the DIRBE-corrected IRAS 100~$\mu$m map \citep{schlegel_maps_1998}.  This infrared emission map is intended to approximate dust mass, which we can convert to total hydrogen column depth by assuming an average gas to dust ratio. The equivalent hydrogen column depth is obtained by multiplying by a factor of $1.44\times10^{20}$~cm$^{-2}$~MJy$^{-1}$, as outlined in \cite{snowden_rosat_1997}.  This approach is favored over other tracers of hydrogen column depth (e.g. 21~cm emission) since it accounts for both atomic and molecular gas---the latter of which can contribute significantly at low galactic latitudes.  Transmission was calculated assuming solar abundances from \cite{anders_solar-system_1982} and cross sections from \cite{balucinska-church_photoelectric_1992}.  The resulting average interstellar transmission for each target area is shown in Figure~\ref{fig:phabs}.

All four spectra were fit with a minimum of four model components:  1) an unabsorbed APEC plasma representing emission from the Local Hot Bubble (LHB); 2) an unabsorbed VACX component representing SWCX due to slow solar wind; 3) an absorbed APEC plasma representing emission from the hot Galactic halo; and 4) an absorbed double broken power law representing the Cosmic X-ray Background (CXRB) due to unresolved extragalactic Active Galactic Nuclei (AGN).  For the low latitude observations, an additional absorbed power law component was included to represent the Crab nebula, which was within the field of view.  The low latitude observations also required an additional thermal component at $\sim$0.9~keV, interpreted as a contribution from dwarf~M (dM) stars and modelled by an absorbed, two-temperature APEC plasma.  Further details of this model and its motivations are presented in Section~\ref{sec:stars}.  Lastly, the spectrum from Mission 27.041 required an additional unabsorbed VACX charge exchange component representing SWCX due to fast solar wind.  This component differs from the slow solar wind primarily in its ion temperature, but also differs slightly in elemental abundances.

\begin{figure*}
\gridline{\fig{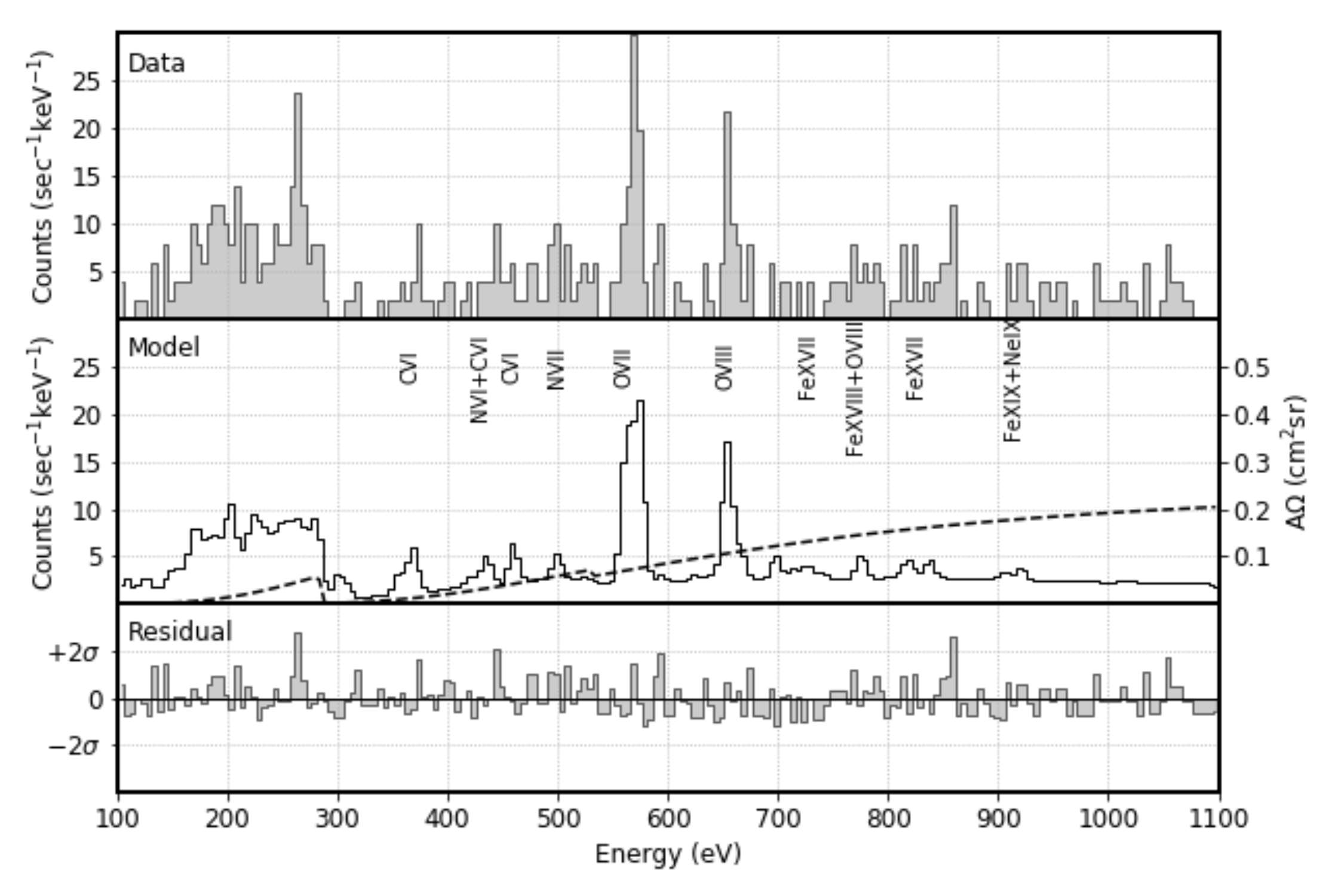}{0.7\textwidth}{27.041}}\label{fig:flt3Spec}
\gridline{\fig{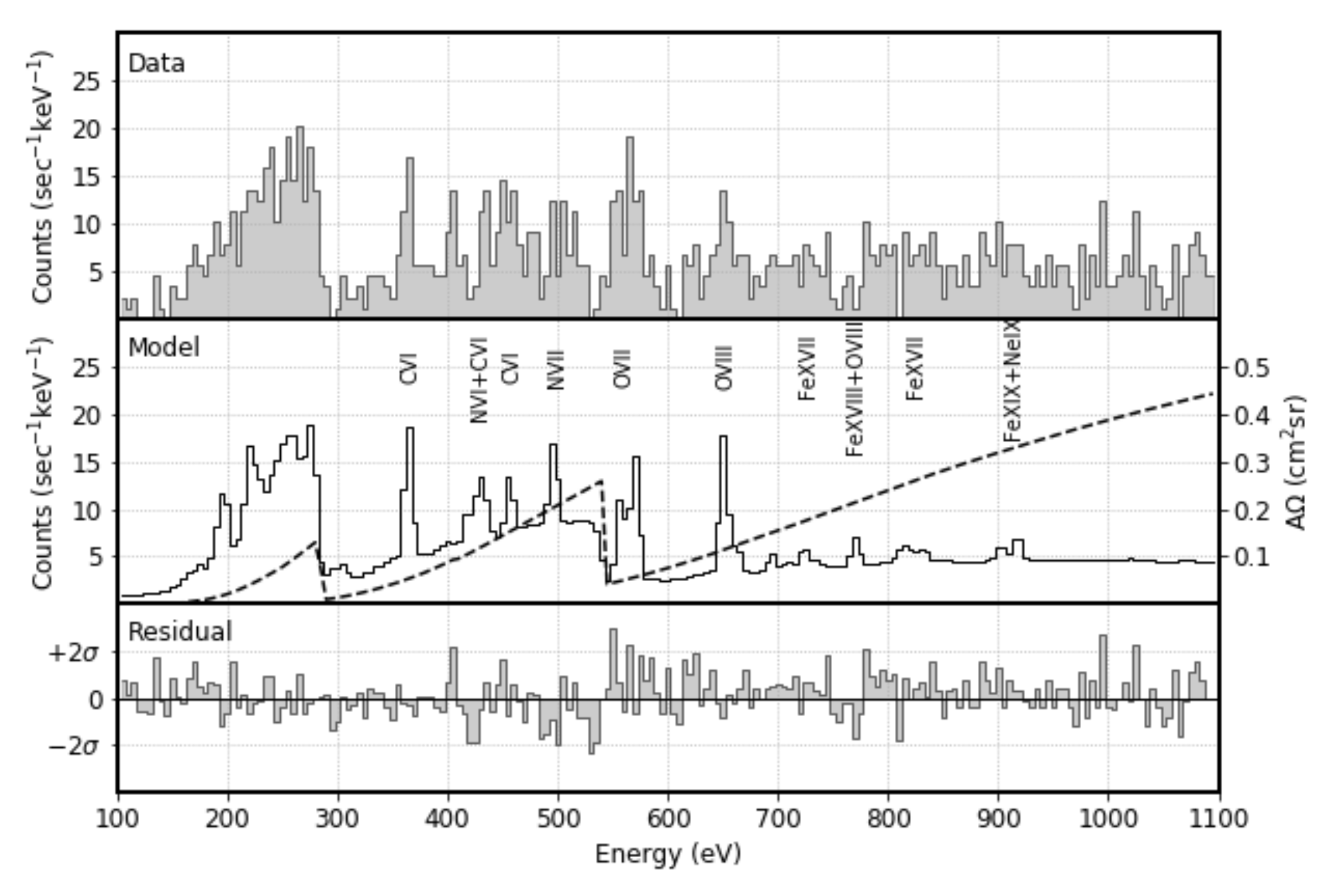}{0.7\textwidth}{36.223}}\label{fig:flt4Spec}
\caption{High Latitude Spectra. Data are shown in the top panel of each figure with the model from Table~\ref{ta:hiModel} shown in the middle panel. The dashed line represents the energy-dependent effective area.  Residuals are shown in the bottom panel in units of sigma, estimated from the model. All histograms have 5~eV bins.}
\end{figure*}

\begin{figure*}
\gridline{\fig{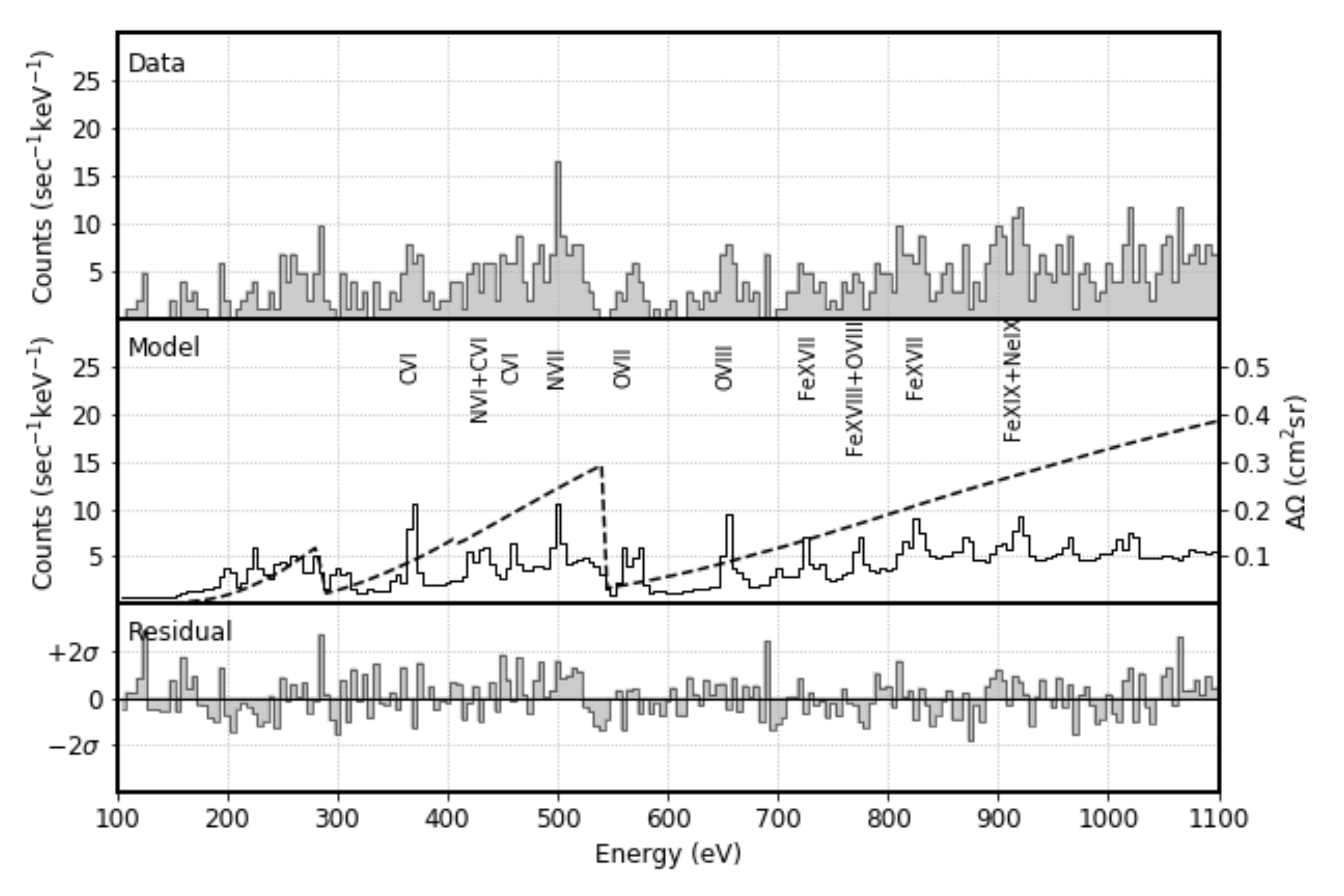}{0.7\textwidth}{36.264}}\label{fig:flt5Spec}
\gridline{\fig{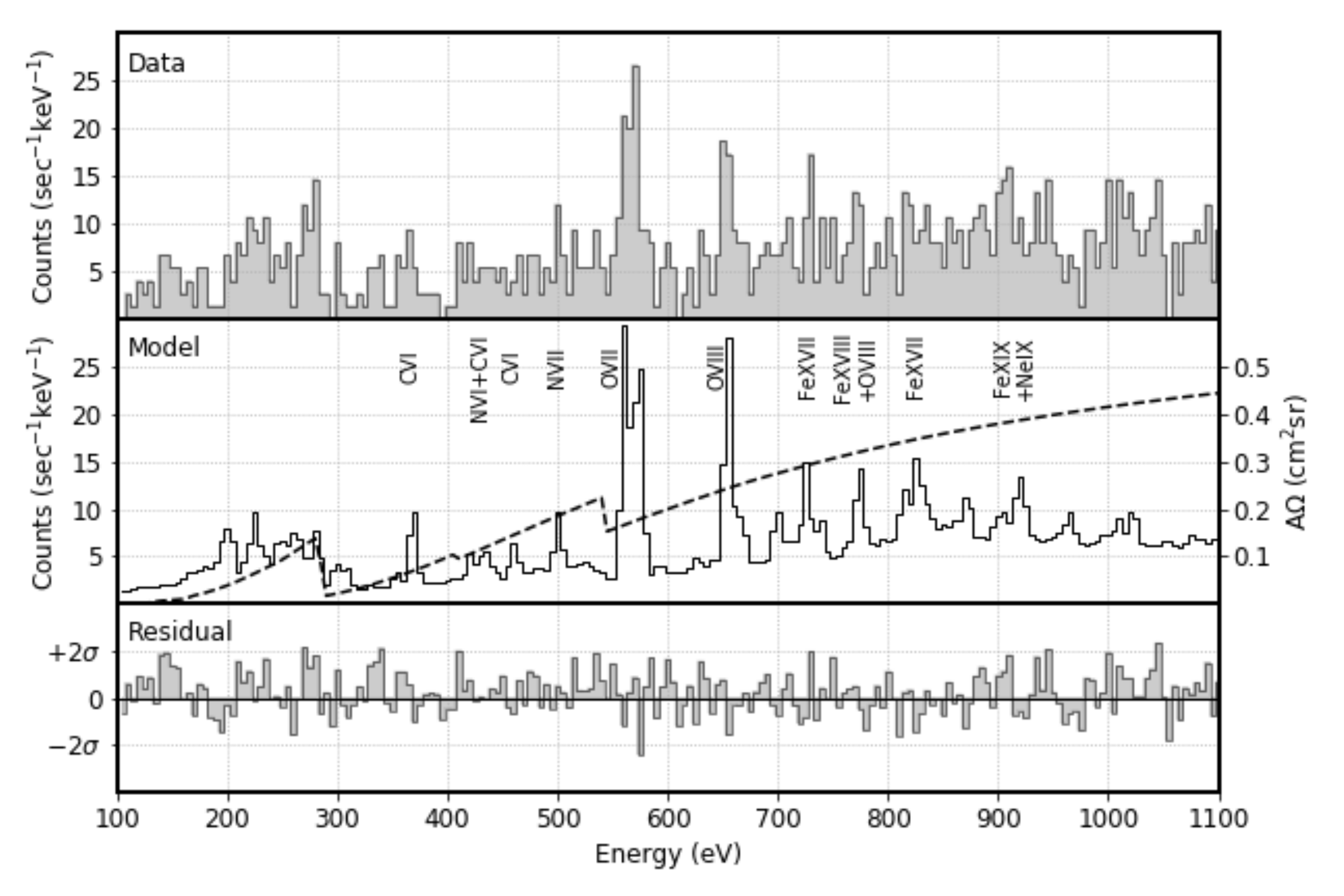}{0.7\textwidth}{36.294}}\label{fig:flt6Spec}
\caption{Low Latitude Spectra. Data are shown in the top panel of each figure with the model from Table~\ref{ta:loModel} shown in the middle panel. The dashed line represents the energy-dependent effective area.  Residuals are shown in the bottom panel in units of sigma, estimated from the model. All histograms have 5~eV bins.}
\end{figure*}

\begin{figure*}
\gridline{\fig{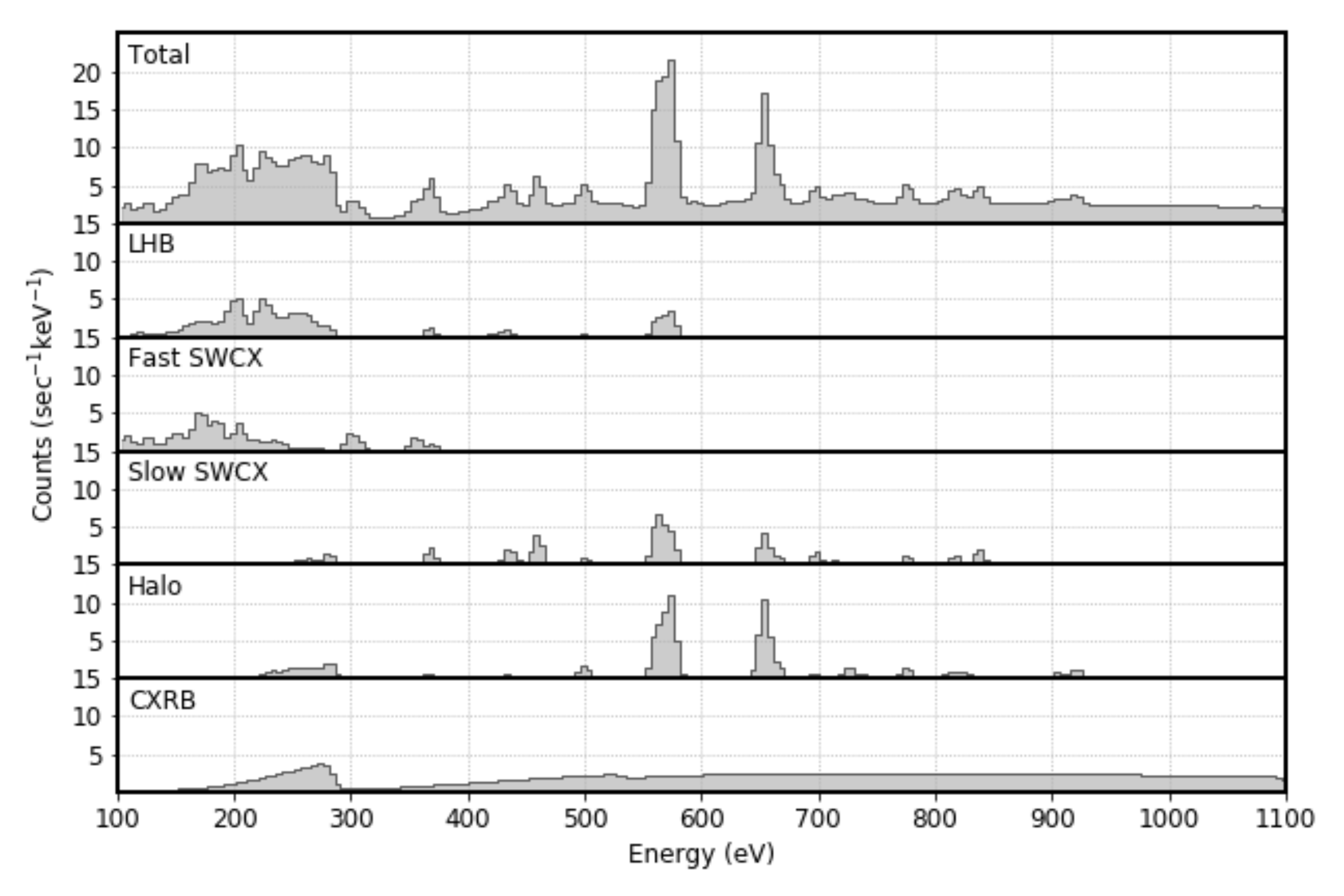}{0.7\textwidth}{27.041}}\label{fig:flt3Model}
\gridline{\fig{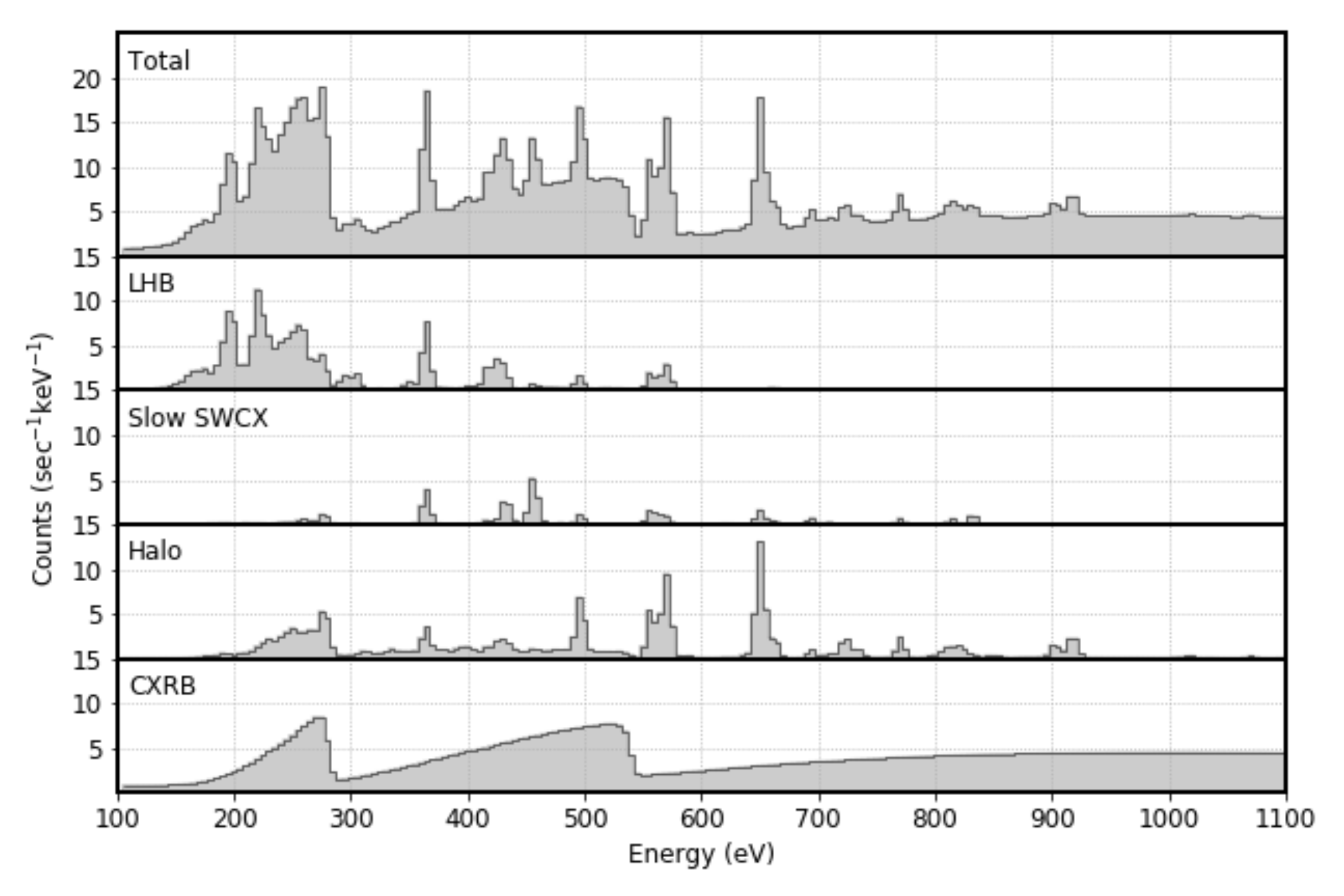}{0.7\textwidth}{36.223}}\label{fig:flt4Model}
\caption{\label{fig:hiLatMod} High latitude model components (Table~\ref{ta:hiModel}). Non~\mbox{X-ray} background is negligible in all cases.}
\end{figure*}

\begin{figure*}
\gridline{\fig{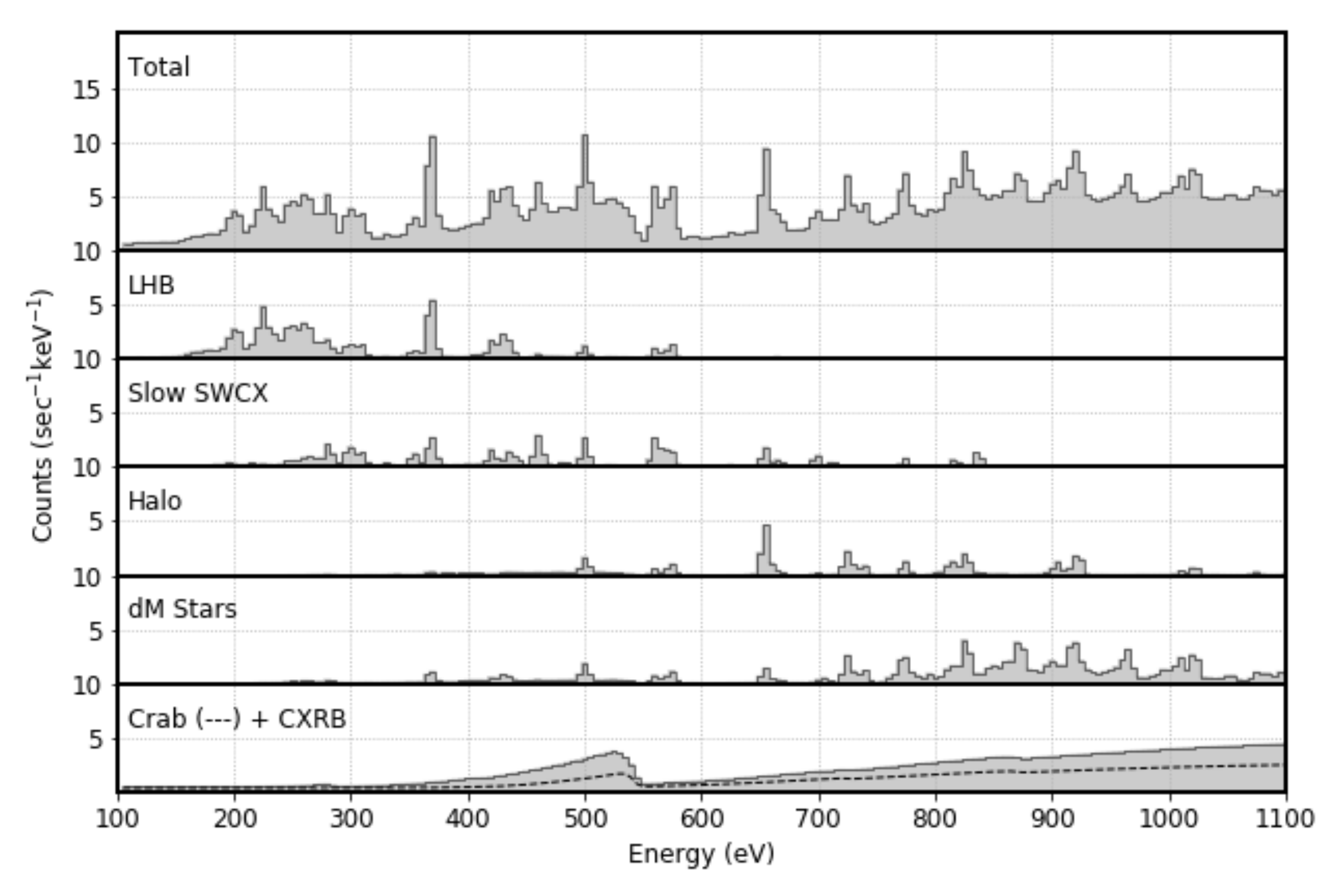}{0.7\textwidth}{36.264}}\label{fig:flt5Model3}
\gridline{\fig{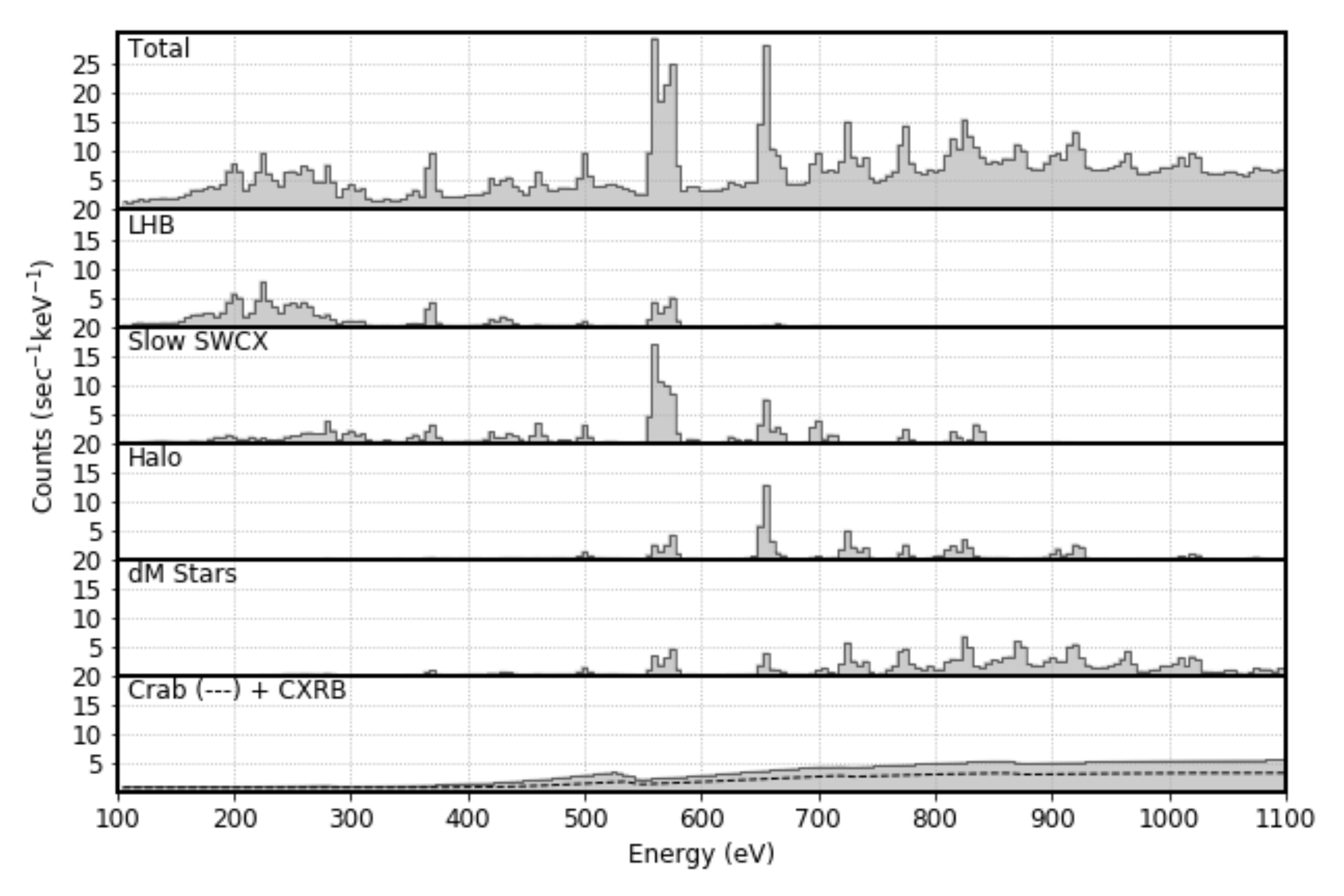}{0.7\textwidth}{36.294}}\label{fig:flt6Model3}
\caption{\label{fig:loLatMod} Low latitude model components (Table~\ref{ta:loModel}). Non~\mbox{X-ray} background is negligible in all cases.}
\end{figure*}

A summary of model components and parameters are listed in Tables~\ref{ta:hiModel} and \ref{ta:loModel}.  Repeat observations of the same target were fit simultaneously with common model parameters, with the exception of the SWCX components, which have free normalizations for each observation.  Parameters for the CXRB and Crab were not fit, and instead were fixed to values from the literature.  Figures~\ref{fig:flt3Spec} through \ref{fig:flt6Spec} show each of the four spectra and corresponding models folded through the instrumental response, and individual model components are depicted in Figures~\ref{fig:hiLatMod} and \ref{fig:loLatMod}.

\begin{table*} 
\caption{Model parameters for simultaneous fit of high latitude observations.  All uncertainties are reported at 90\% confidence level. In instances where a component or parameter is unique to a single spectrum, the values are reported in the column of the corresponding spectrum.}
\label{ta:hiModel}
	\begin{center}
	{\footnotesize
	\begin{tabular}{l c c c c}
	\hline
	&  & & \multicolumn{2}{c}{Value} \\
	Component & Model & Parameter & 27.041 & 36.223\\
	\hline
	Local Hot & APEC$^{\rm a}$ & $kT$ (keV) & \multicolumn{2}{c}{0.093$^{+0.010}_{-0.019}$} \\
	Bubble & & Normalization$^{\rm b}$  & \multicolumn{2}{c}{8.09$^{+1.92}_{-1.79}$} \\
	\hline
	Fast  SWCX & VACX$^{\rm c}$ & $kT$ (keV) & 0.055$^{+0.011}_{-0.006}$ &\\
	& & Normalization$^{\rm d}$ & 4.17$^{+3.54}_{-2.25}$ &\\
	\hline
	Slow  SWCX & VACX$^{\rm c}$ & $kT$ (keV) & \multicolumn{2}{c}{0.184$^{+0.042}_{-0.045}$} \\
	& & Normalization$^{\rm d}$ & 0.073$^{+0.059}_{-0.042}$ & 0.023$^{+0.019}_{-0.016}$ \\
	\hline
	Hot Halo & Absorbed & $kT$ (keV) & \multicolumn{2}{c}{0.189$^{+0.026}_{-0.018}$} \\
	& APEC$^{\rm a,e}$ & Normalization$^{\rm b}$ & \multicolumn{2}{c}{4.99$^{+2.04}_{-1.75}$}  \\
	\hline
	CXRB & Absorbed Double &  Photon Index 1 & \multicolumn{2}{c}{1.54} \\
	& Broken Power Law$^{\rm e,f}$ & ($E< E_{break}$) & & \\
	& & Normalization 1$^{\rm g}$ & \multicolumn{2}{c}{5.7} \\
	& & Photon Index 2 & \multicolumn{2}{c}{1.96} \\
	& & ($E< E_{break}$) & & \\
	& & Normalization 2$^{\rm g}$ & \multicolumn{2}{c}{4.9} \\
	& & $E_{break}$ (keV) & \multicolumn{2}{c}{1.2} \\
	& & Photon Index & \multicolumn{2}{c}{1.4} \\
	& & ($E> E_{break}$) & & \\
	\hline
	$\chi^{2}$/dof $^{\rm h}$ & & & \multicolumn{2}{c}{997.67/788 = 1.266} \\
	\hline
	\multicolumn{5}{l}{$^{\rm a}$ Solar abundances from \cite{anders_abundances_1989}.}\\
	\multicolumn{5}{l}{$^{\rm b}$ Units of $10^{14}$ cm$^{-5}$ sr$^{-1}$.} \\
	\multicolumn{5}{l}{$^{\rm c}$ Relative abundances from Table~1 in \cite{von_steiger_composition_2000}.}\\
	\multicolumn{5}{l}{$^{\rm d}$ Parameter has no intrinsic physical unit.  See text for discussion.} \\
	\multicolumn{5}{l}{$^{\rm e}$ Transmission-weighted average interstellar absorption (Figure~\ref{fig:phabs}).}\\
	\multicolumn{5}{l}{$^{\rm f}$ Double power law model given by \cite{smith_suzaku_2007}}\\
	\multicolumn{5}{l}{$^{\rm g}$ Units of photons keV$^{-1}$ s$^{-1}$ cm$^{-2}$ sr$^{-1}$ at 1~keV.}\\
	\multicolumn{5}{l}{$^{\rm h}$ Pearson's $\chi^{2}$ statistic \citep{f.r.s_x._1900}}\\
	\end{tabular}
	}
	\end{center}
\end{table*}

\begin{table*} 
\caption{Same as Table~\ref{ta:hiModel} except for low latitude observations.}
\label{ta:loModel}
	\begin{center}
	{\footnotesize
	\begin{tabular}{l c c c c}
	\hline
	&  &  & \multicolumn{2}{c}{Value} \\
	Component & Model & Parameter & 36.264 & 36.294\\
	\hline
	Local Hot & APEC & $kT$ (keV) & \multicolumn{2}{c}{0.099 (fixed)} \\
	Bubble & & Normalization & \multicolumn{2}{c}{3.15$^{+0.72}_{-0.69}$} \\
	\hline
	Slow  SWCX & VACX & $kT_{C}$ (keV) & \multicolumn{2}{c}{0.088$^{+0.011}_{-0.020}$} \\
	& & $kT$ (keV) & \multicolumn{2}{c}{0.176$^{+0.016}_{-0.020}$} \\
	& & Normalization & 0.047$^{+0.028}_{-0.023}$ & 0.069$^{+0.033}_{-0.027}$ \\
	\hline
	Hot Halo & Absorbed & $kT$ (keV) & \multicolumn{2}{c}{0.245$^{+0.069}_{-0.059}$} \\
	& APEC & Normalization & \multicolumn{2}{c}{4.92$^{+4.28}_{-3.08}$}  \\	
	\hline
	dM Stars & Absorbed & Normalization$^{\rm a}$ & \multicolumn{2}{c}{1.00$^{+0.27}_{-0.26}$} \\
	& Two-Temperature & & & \\
	& APEC & & & \\
	\hline
	CXRB & Absorbed Double &  Photon Index 1 & \multicolumn{2}{c}{1.54} \\
	& Broken Power Law & ($E< E_{break}$) & & \\
	& & Normalization 1 & \multicolumn{2}{c}{5.7} \\
	& & Photon Index 2 & \multicolumn{2}{c}{1.96} \\
	& & ($E< E_{break}$) & & \\
	& & Normalization 2 & \multicolumn{2}{c}{4.9} \\
	& & $E_{break}$ (keV) & \multicolumn{2}{c}{1.2} \\
	& & Photon Index & \multicolumn{2}{c}{1.4} \\
	& & ($E> E_{break}$) & & \\
	\hline
	Crab & Absorbed & $N_{H}$ (10$^{22}$ cm$^{-2}$)& \multicolumn{2}{c}{0.43} \\
	& Power Law$^{\rm b}$ & Photon Index & \multicolumn{2}{c}{2.04} \\
	& & Normalization$^{\rm c}$ & \multicolumn{2}{c}{8.8} \\
	\hline
	$\chi^{2}$/dof & & & \multicolumn{2}{c}{1096.66/788 = 1.392} \\
	\hline
	\multicolumn{5}{l}{$^{\rm a}$ Dimensionless. See text for description of model and normalization parameter.}\\
	\multicolumn{5}{l}{$^{\rm b}$ Absorbed power law given by Table~1 in \cite{weisskopf_calibrations_2010}.}\\
	\multicolumn{5}{l}{$^{\rm c}$ Units of photons keV$^{-1}$ s$^{-1}$ cm$^{-2}$ at 1~keV.}\\
	\end{tabular}
	}
	\end{center}
\end{table*}

\section{Discussion}\label{sec:discussion}

\subsection{Solar Wind Charge Exchange}\label{sec:cx}
The VACX model used to generate SWCX spectra is only approximate, since it lacks any dependence on the solar wind velocity.  Laboratory measurements of charge exchange emission spectra have demonstrated that the relative strength of different emission lines changes significantly over the range of solar wind velocities \citep[e.g.][]{defay_x-ray_2013}. There are also large
uncertainties and variabilities in the metal ion content of the solar wind.  Nevertheless, the two SWCX components used here improve the fits and help resolve differences between repeat observations of the same target---especially at high latitudes.  With the exception of Mission 27.041, the relative contribution tends to be less than estimates for heliospheric SWCX in the RASS R12 maps (Table~\ref{ta:swcxFlux}). This difference is likely responsible for the discrepancy in Table~\ref{ta:check} at low energies.  As previously mentioned, 27.041 is the only observation to require a significant contribution from fast solar wind. For the simultaneous fit to the high latitude observations, the addition of the fast solar wind component to the 27.041 spectral model improves the goodness of fit by $\Delta \chi^{2} / \Delta dof = 83.99 / 2$. Coincidentally, this is also the only observation to exhibit strong excess emission at 263~eV (See Figure~\ref{fig:flt3Spec}).  Perhaps this feature may be attributed to L-shell emission from solar wind ions S{\scriptsize ~IX} or Si{\scriptsize ~X}, but which is not included in this single-temperature model for the fast solar wind.

\begin{table*} 
\caption{Combined contribution of fast and slow SWCX model components in Tables~\ref{ta:hiModel} and \ref{ta:loModel} to the ROSAT energy bands R1 and R2, compared to estimates for the RASS R1 and R2 maps given by \cite{uprety_solar_2016}. All uncertainties are reported at 90\% confidence level.}
\label{ta:swcxFlux}
	\begin{center}
	{\footnotesize
	\begin{tabular}{l l c c}
	\hline
	& & \multicolumn{2}{c}{SWCX Flux (RU$^{\rm a}$)} \\
	Target & Mission & R1 & R2 \\
	\hline
	High Latitude & 27.041 & 489$^{+414}_{-265}$ & 234$^{+197}_{-127}$  \\
	& 36.223 & 22$^{+19}_{-16}$ & 12$^{+11}_{-9}$ \\
	& RASS & 70 & 110 \\
	\hline
	Low Latitude & 36.264 & 22$^{+13}_{-11}$ & 18$^{+10}_{-9}$  \\
	& 36.294 & 28$^{+13}_{-11}$ & 25$^{+12}_{-10}$ \\
	& RASS & 55 & 81 \\
	\hline
	\multicolumn{4}{l}{$^{\rm a}$1 RU $\equiv$ 10$^{-6}$ counts s$^{-1}$ arcmin$^{-2}$} \\
	\end{tabular}
	}
	\end{center}
\end{table*}

A more robust measurement of the SWCX contribution can be made by analyzing individual emission lines, made possible by the high spectral resolution of these observations.  After fitting a global model to each spectrum, the intensities of individual lines are measured by setting the abundance for a single element to zero in each of the model components and adding $\delta$-functions for each of the strong lines of that element.  Fluxes measured in this manner for select emission lines are presented in Table~\ref{ta:lineID}, with ratios of those lines given in Table~\ref{ta:ratios}.  Of particular interest are lines from C{\scriptsize ~VI} and O{\scriptsize ~VII}, since they are relatively bright, are isolated from other lines, and can be used to unambiguously differentiate between thermal emission and SWCX.  

\begin{table*} 
\caption{Line identification and fluxes. F, I, and R refer to the forbidden, intercombination, and resonance lines due to the fine structure splitting of the He-like K-$\alpha$ triplet.  The O{\scriptsize ~VII} K-$\alpha$ fine structure splitting could not be resolved for mission 27.041, though a total flux of 4.74$^{+1.52}_{-1.41}$~LU is observed for the triplet. All uncertainties and limits are reported at 90\% confidence level.}
\label{ta:lineID}
	\begin{center}
	{\footnotesize
	\begin{tabular}{l c c c c c}
	\hline
	& Energy & \multicolumn{4}{c}{Flux (LU$^{\rm a}$)}\\
	Emission Line & (eV) & 27.041 & 36.223 & 36.264 & 36.294\\
	\hline
	C{\tiny ~VI} Ly-$\alpha$ & 368 & 5.88$^{+5.15}_{-3.55}$ & 2.02$^{+1.03}_{-0.84}$ & 0.77$^{+0.46}_{-0.36}$ & 0.86$^{+0.68}_{-0.49}$ \\
	C{\tiny ~VI} Ly-$\gamma$ & 459 & $<$2.02 & 0.38$^{+0.37}_{-0.29}$ & 0.24$^{+0.21}_{-0.16}$ & $<$0.36 \\
	O{\tiny ~VII} K-$\alpha$ F & 561 & - & 1.88$^{+1.20}_{-0.93}$ & 0.69$^{+0.82}_{-0.58}$ & 1.12$^{+0.48}_{-0.40}$ \\
	O{\tiny ~VII} K-$\alpha$ I & 569 & - & $<$2.07 & $<$1.37 & 0.59$^{+0.49}_{-0.41}$ \\
	O{\tiny ~VII} K-$\alpha$ R & 575 & - & 2.37$^{+1.18}_{-0.96}$ & 0.59$^{+0.75}_{-0.50}$ & 0.62$^{+0.44}_{-0.35}$ \\
	O{\tiny ~VIII} Ly-$\alpha$ & 654 & 1.48$^{+0.79}_{-0.61}$ & 0.91$^{+0.48}_{-0.39}$ & 0.74$^{+0.47}_{-0.35}$ & 0.69$^{+0.31}_{-0.25}$ \\
	\hline
	\multicolumn{6}{l}{$^{\rm a}$1 LU $\equiv$ 1 photon~sec$^{-1}$~cm$^{-2}$~sr$^{-1}$}\\
	\end{tabular}
	}
	\end{center}
\end{table*}

\begin{table*} 
\caption{Ratios of lines identified in Table~\ref{ta:lineID}.  Statistical uncertainties estimated from the model are given at the 1$\sigma$ level.}
\label{ta:ratios}
	\begin{center}
	{\footnotesize
	\begin{tabular}{l c c c c c c}
	\hline
	& & & & & Thermal & SWCX\\
	Ratio & 27.041 & 36.223 & 36.264 & 36.294 & Prediction$^{\rm a}$ & Prediction$^{\rm b}$\\
	\hline
	C{\tiny ~VI} Ly-$\gamma$/Ly-$\alpha$ & 0.13$\pm$0.09 & 0.19$\pm$0.07 & 0.31$\pm$0.13 & 0.10$\pm$0.08 & 0.04 & 0.5\\
	O{\tiny ~VII} (F$+$I)/R$^{\rm c}$ & - & 1.19$\pm$0.33 & 2.03$\pm$1.07 & 2.76$\pm$0.79 & 0.97 & 6.0\\
	\hline
	\multicolumn{7}{l}{$^{\rm a}$Based on APEC components in spectral model.} \\
	\multicolumn{7}{l}{\parbox[t]{5in}{$^{\rm b}$Line ratios for SWCX depend on both velocity and electron donor. Reported values are upper limits to conservatively estimate minimum possible SWCX contribution.}} \\
	\multicolumn{7}{l}{$^{\rm c}$Also known as the $G$-ratio.} \\
	\end{tabular}
	}
	\end{center}
\end{table*}

As noted by a previous analysis of Mission 36.223, the Ly-$\gamma$/Ly-$\alpha$ ratio of H-like C{\scriptsize ~VI} and the $G$-ratio of He-like O{\scriptsize ~VII} can be used to constrain the allowed thermal contribution to the observed emission from these respective ions \citep{crowder_observed_2012}.  Although the ratios for SWCX spectra depend on both collision velocity and electron donor (e.g hydrogen versus helium), upper limits can be used to conservatively estimate the minimum possible contribution from SWCX \citep{greenwood_experimental_2001,koutroumpa_etude_2007,defay_x-ray_2013,fogle_x-ray-emission_2014}.  For an observed line ratio $R_{obs}$ and predicted ratios for thermal and SWCX emission, $R_{therm}$ and $R_{swcx}$, the inferred fraction due to thermal emission, $F_{therm}$, is given by Equation~\ref{eq:thermFrac}.

\begin{equation}
\label{eq:thermFrac}
F_{therm} = \frac{(1+R_{therm})(R_{swcx}-R_{obs})}{(1+R_{swcx})(R_{swcx}-R_{therm})}
\end{equation}

\noindent Table~\ref{ta:fracTherm} presents the results of this analysis.  In agreement with the previous analysis, we find that thermal emission likely dominates the observed emission for both C{\scriptsize ~VI} and O{\scriptsize ~VII} at high latitudes, though the statistical uncertainties allow for a non-negligible contribution from SWCX.  

\begin{table*} 
\caption{Relative thermal contribution to observed emission lines from select ions, derived from line ratios given in Table~\ref{ta:ratios} using Equation~\ref{eq:thermFrac}.}  Statistical uncertainties estimated from the model are given at the 1$\sigma$ level.
\label{ta:fracTherm}
	\begin{center}
	{\footnotesize
	\begin{tabular}{l c c c c}
	\hline
	& \multicolumn{4}{c}{Thermal Fraction} \\
	\multicolumn{1}{c}{Ion} & 27.041 & 36.223 & 36.264 & 36.294 \\
	\hline
	C{\tiny ~VI} & 0.75$^{+0.25}_{-0.21}$ & 0.59$^{+0.18}_{-0.16}$ & 0.33$^{+0.29}_{-0.23}$ & 0.81$^{+0.19}_{-0.21}$ \\
	O{\tiny ~VII} & - & 0.86$^{+0.14}_{-0.17}$ & 0.51$^{+0.49}_{-0.24}$ & 0.34$^{+0.20}_{-0.13}$ \\
	\hline
	\end{tabular}
	}
	\end{center}
\end{table*}

At low latitudes we find that SWCX is likely responsible for at least half of the observed O{\scriptsize ~VII} emission, though this seems to be due to a reduction in thermal emission compared to the high latitude observations, rather than an increase in SWCX.  At all latitudes, we find that SWCX accounts for $\sim$1~LU (1 LU $\equiv$ 1 photon~sec$^{-1}$~cm$^{-2}$~sr$^{-1}$) of O{\scriptsize ~VII}~K$\alpha$ emission.  SWCX may also be dominating C{\scriptsize ~VI} emission in the Mission 36.264 observation, though this not statistically very significant.  Initial fits to the low latitude spectra using the VACX model failed to predict the strong relative observed contribution to O{\scriptsize ~VII}.  In order to reproduce the O{\scriptsize ~VII} flux inferred from the $G$-ratio, the ion temperature of carbon in the VACX model ($kT_{C}$ in Table~\ref{ta:loModel}) was allowed to vary independently of the temperature of the other ions.  

Archival data available through OMNIWeb\footnote{\url{http://omniweb.gsfc.nasa.gov/}} was used to get solar wind proton fluxes during each observation, time-shifted to Earth. The fluxes are $<3\times10^{8}$~cm$^{-2}$s$^{-1}$ during each observation, which is below the level that \cite{yoshino_energy_2009} finds to be associated with significant contributions from geocoronal SWCX in Suzaku observations.  Moreover, the near midnight zenith viewing geometry of sounding rocket observations   makes them less susceptible to contamination from geocoronal SWCX, which originates most strongly from the Earth's subsolar magnetosheath \citep{robertson_spatial_2003}.  For these reasons, it is more likely that SWCX emission observed here is of heliospheric orign.

\subsection{Hot Galactic Halo}\label{sec:halo}
A single temperature halo model was used for this analysis, with a best-fit temperature of $\sim$0.2~keV at both low and high latitudes, assuming solar abundances.  This is consistent with other observations which fit single temperature halos with $kT\sim0.2$~keV \citep{yoshino_energy_2009}. However, previous observations have also established that the thermal and spatial distribution of the halo is complex \citep[e.g.][]{kuntz_deconstructing_2000, lei_determining_2009, yao_x-ray_2009, liu_evidence_2016}, and a single temperature model is therefore likely overly simplistic.  One should be careful not to ascribe too much significance to any physical interpretation. 

There is good evidence from cloud shadowing, small-scale absorption studies, and observations at lower energies that bright patches of $\sim$0.1~keV emission in the Galactic halo contribute substantially to our high-latitude  observations \citep{burrows_soft_1991, burrows_probing_1993, snowden_progress_1998, kuntz_deconstructing_2000, bellm_origins_2005}.  However, an attempt to add an additional low-temperature component to our model resulted in no improvement to the fit.  This is not surprising, given that these observations lack the spatial resolution used by other analyses to separate this component from the local plasma at approximately the same temperature.  Therefore our LHB fluxes necessarily include this low-temperature halo contribution, as will be discussed further in Section~\ref{sec:lhb}.

\subsection{The Local Hot Bubble}\label{sec:lhb}
At both high and low latitudes, we fit a local plasma temperature of $\sim$0.1~keV, using solar abundances.  For reference, \cite{snowden_progress_1998} reports 0.10~keV from an analysis of the RASS data. More recently, \cite{liu_structure_2017} reanalyzed the RASS data taking contributions from SWCX into account and reports a temperature of 0.097~keV.  \cite{smith_resolving_2014} modeled SWCX contributions to the DXS hot interstellar medium spectrum \citep{sanders_spectra_2001} with the same VACX model used for this analysis and reports a LHB temperature of 0.097~keV with depleted abundances.  

With the exception of the DXS observation, the data presented here have significantly better energy resolution than previous observations. Moreover, the aforementioned analyses rely primarily on data below the carbon edge at 282~eV.  Observations that include higher energy data including O{\scriptsize ~VII} K$\alpha$ emission at $\sim$570~eV tend to fit higher temperatures that are inconsistent with lower energy observations, e.g. \cite{yoshino_energy_2009}.  For the observations analyzed here, we find that the fit temperature of the LHB is very sensitive to the intensity of O{\scriptsize ~VII}.  If included in the fit, we derive a low latitude LHB temperature that is inconsistent with the high latitude observations.  However, if the energy range is restricted to $<$500~eV during the fit, then the 90\% confidence temperature ranges agree at both latitudes (See Tables~\ref{ta:hiModel} and \ref{ta:loModel}).  As mentioned in Section~\ref{sec:cx}, O{\scriptsize ~VII} is likely dominated by SWCX at low latitudes and this contribution is difficult to estimate with sufficient accuracy.  Therefore, in this analysis the LHB temperature at low latitudes is fixed to the value derived from the low energy spectrum before fitting the entire 100--1100~eV range.  

As mentioned in Section~\ref{sec:halo}, at high latitudes we are unable to separate the local plasma from the low temperature component of the halo, which affects our derived values for emission measure, as well as possibly temperature. \cite{liu_structure_2017} combined SWCX maps from \cite{uprety_solar_2016} with data from \cite{snowden_progress_1998} to estimate the local and distant astrophysical contributions to the R12 band.  In Table~\ref{ta:R12Flux}, we compare the results of our spectral model fits to that work.  At low latitudes, where the distant R12 emission is strongly absorbed, we find good agreement.  At high latitudes, however, we overestimate the ratio of local to distant flux as we would expect without a low temperature halo component.  Nevertheless, at both latitudes our total derived flux from astrophysical sources is remarkably consistent with \cite{liu_structure_2017}.  This provides a strong indication that the discrepancies seen at low energies in Table~\ref{ta:check} do indeed arise from differences in heliospheric SWCX contribution.

\begin{table*} 
\caption{Comparison of local and distant contributions to ROSAT R12 band fluxes after correcting for contribution from SWCX.  SWCX-corrected values for the RASS given by \cite{liu_structure_2017}. Statistical uncertainties estimated from the model are given at the 1$\sigma$ level.}
\label{ta:R12Flux}
	\begin{center}
	{\footnotesize
	\begin{tabular}{l l c c c}
	\hline
	& & \multicolumn{3}{c}{Astrophysical R12 Flux (RU)} \\
	Target & Instrument & Local & Distant & Total \\
	\hline
	High Latitude & XQC & 495$\pm$85 & 380$\pm$72 & 827$\pm$89 \\
	& RASS & 383 & 448 & 831\\
	\hline
	Low Latitude & XQC & 228$\pm$44 & 47$\pm$36 & 272$\pm$46\\
	& RASS & 247 & 17 & 264 \\
	\hline
	\end{tabular}
	}
	\end{center}
\end{table*}

\subsection{Stellar Contribution}\label{sec:stars}
At low latitudes, we observe excess emission around 0.9~keV arising from lines of  Ne{\scriptsize ~IX} and Fe{\scriptsize ~XVII--XIX}, which suggests the existence of an additional thermal component of higher temperature than those previously discussed.  While evidence of 0.7--1~keV diffuse gas has been observed in outflows toward the Galactic center \citep[e.g.][]{ponti_x-ray_2019} and in the direction of the North Polar Spur \citep{das_discovery_2019}, it is difficult to confine a diffuse source of this temperature to low Galactic latitudes.  Therefore, we investigated stars as a possible source of the emission, which have long been proposed to explain the excess 3/4~keV flux throughout the Galactic plane \citep[e.g.][]{rosner_stellar_1981,kuntz_contribution_2001,masui_nature_2009, gupta_contribution_2009}. Following \cite{kuntz_contribution_2001} and \cite{masui_nature_2009}, we constructed a spectral emission model of dwarf~M stars (dM stars), taking into account Galactic absorption and spatial star distribution. Stellar spectra were approximated as a two temperature plasma with \mbox{T$_{\rm C}=0.138$~keV} and \mbox{T$_{\rm H}=0.78$~keV}, based on ROSAT observations of nearby dM stars \citep{giampapa_coronae_1996}. Parameters detailing the spatial emission measure distribution of each temperature component are given in Table~\ref{ta:starMod}, normalized to \mbox{$4.13\times10^{14}$~cm$^{-5}$~sr$^{-1}$~kpc$^{-1}$} in the plane of the Galaxy at the solar radius. Absorption was calculated from Equations~1 and 2 in \cite{ferriere_interstellar_2001}. We then fit low latitude observations with this stellar model scaled by a single free normalization parameter. The best-fit value of this parameter is consistent with unity (1.00$^{+0.27}_{-0.26}$, see Table~\ref{ta:loModel}), providing quantitative agreement between the observed excess flux and model prediction. Moreover, the addition of this component provides a significant improvement to the goodness of fit ($\Delta \chi^{2} / \Delta dof = 103.7 / 1$). Since the scale height of the young ($<$0.15 Gyr) dM stars responsible for 71\% of the \mbox{X-ray} emission in our model is even smaller than the gas scale height, the stellar \mbox{X-ray} contribution is expected to drop sharply with increasing Galactic latitude. Trying to fit this model to our high latitude observations indicates an even steeper drop-off, with a 2$\sigma$ upper limit of 1.05 for the normalization.

\begin{table*} 
\caption{Spatial density distribution parameters used in spectral emission model for combined early- and late-type dM stars.}
\label{ta:starMod}
	\begin{center}
	{\footnotesize
	\begin{tabular}{c c c c c}
	\hline
	 &  &  & & Midplane EM Density$^{c}$ \\
	Age (Gyr) & $K^{a}_{+}$ (pc) & $K^{a}_{-}$ (pc) & $c^{a,b}$ & ($10^{14}$~cm$^{-5}$~sr$^{-1}$~kpc$^{-1}$) \\
	\hline
	0--0.15 & 5000 & 3000 & 0.0140 & 2.93 \\
	0.15--1 & 2226 & 494.4 & 0.0279 & 0.85 \\
	1--2 & 2226 & 494.4 & 0.0457 & 0.07 \\
	2--3 & 2226 & 494.4 & 0.0662 & 0.05 \\
	3--5 & 2226 & 494.4 & 0.0867 & 0.07 \\
	5--10  & 2226 & 494.4 & 0.0958 & 0.16 \\
	\hline
	Total & & & & 4.13\\
	\hline
	\multicolumn{5}{l}{$^{a}$Scale parameters for (1) and (2) of Appendix A in \cite{bienayme_mass_1987}.} \\
	\multicolumn{5}{l}{$^{b}$Dimensionless ellipticity from Notes of Table 2 in \cite{kuntz_contribution_2001}.} \\
	\multicolumn{5}{l}{$^{c}$Adapted from Table 2 in \cite{kuntz_contribution_2001} and Table 3 in \cite{masui_nature_2009}.} \\
	\end{tabular}
	}
	\end{center}
\end{table*}

\section{Conclusions}\label{sec:conclusion}
Here we presented a combined analysis of the past four XQC sounding rocket observations.  The three most recent observations have all been processed with an overlapped pulse fitting algorithm to improve spectral resolution and reduce processing deadtime compared to conventional processing techniques.  The resulting spectra are consistent with fluxes measured by the RASS, with discrepancies reasonably accounted for by the estimated contributions from SWCX.  Likewise, repeated XQC observations of the same targets can be fit with identical spectral emission models, adjusting for differences in instrumental response and allowing for variable SWCX intensity.  Analyses of individual emission lines of C{\scriptsize ~VI} and O{\scriptsize ~VII} permit a more robust measurement of the SWCX contribution and LHB temperature that is less sensitive to inaccuracies in the spectral models for charge exchange.  The derived temperatures for the LHB and halo are consistent between high and low latitude observations, and compatible with prior values reported in the literature.  Emission around 0.9~keV observed only at low latitudes can be explained by unresolved dM stars.  

These results are not presented as a complete description of the sources of the Galactic DXRB. The spectra from these brief sounding rocket observations (which consist of only two pointings with less than 11 minutes of observing time) have relatively poor statistical precision and average over a large solid angle. Thus a relatively simple model can suffice for these observations---one that neglects complexities that are well-established from other measurements. Examples include the strong case against a single-temperature halo from emission-absorption studies \citep{yao_nonisothermality_2007, yao_x-ray_2009}, convincing evidence for substantial contributions to 1/4~keV emission at high latitudes from 1.0$\times$10$^6$~K emitting regions in a multi-phase halo (e.g. \cite{kuntz_deconstructing_2000}), and the finding that the hot halo is very non-uniform, with 20\% of high latitude on/off cloud observations showing almost no distant O{\scriptsize ~VII} or O{\scriptsize ~VIII} emission and generally very different distributions of emissivity in these two lines \citep{henley_xmm-newtonandsuzakux-ray_2015}. Additionally, the models used for SWCX are quite crude in comparison to thermal emission models. Solar wind ion abundances are highly variable and unlikely to correspond to a single temperature as assumed here, and the line emission ratios are derived from models known to be inaccurate by factors of two and more at solar wind velocities. 

But there is the promise of major improvements in the near future. We have already used this same XQC payload as a detector on the merged beam collision facility at Oak Ridge National Laboratories to make some preliminary measurements of line ratios from charge exchange as a function of velocity that illustrate limitations of current calculations \citep{defay_x-ray_2013, fogle_x-ray-emission_2014, seely_line_2017}. The installation has now been greatly improved and will soon be collecting better data on lines from ion-neutral pairs of astrophysical interest. The eROSITA mission should soon begin a series of eight ROSAT-like all sky surveys with CCD energy resolution. A small orbital mission of an XQC-like instrument could quickly improve on the results presented here by orders of magnitude, and the very high spectral resolution would nicely complement the very high spatial resolution and statistical precision of the eROSITA surveys.

\bigskip

We would like acknowledge the contributions of many graduate and undergraduate students who have contributed to the XQC instrument.  We also thank the Luxel Corporation and Kari Kripps for the fabrication of the IR-blocking filters and support meshes.   We are grateful for the support of the sounding rocket staff at Wallops Flight Facility and White Sands Missile Range. Finally, we thank the anonymous referee for their careful reading of the manuscript and helpful comments. The contributions of the Goddard co-authors were supported by the National Aeronautics and Space Administration under an award issued through its Astrophysics Research and Analysis Program. This work was supported in part by NASA grant NNX16AM31G.




\bibliography{thesis}

\begin{thebibliography}{}
\expandafter\ifx\csname natexlab\endcsname\relax\def\natexlab#1{#1}\fi
\providecommand{\url}[1]{\href{#1}{#1}}
\providecommand{\dodoi}[1]{doi:~\href{http://doi.org/#1}{\nolinkurl{#1}}}
\providecommand{\doeprint}[1]{\href{http://ascl.net/#1}{\nolinkurl{http://ascl.net/#1}}}
\providecommand{\doarXiv}[1]{\href{https://arxiv.org/abs/#1}{\nolinkurl{https://arxiv.org/abs/#1}}}

\bibitem[{Anders \& Ebihara(1982)}]{anders_solar-system_1982}
Anders, E., \& Ebihara, M. 1982, Geochimica et Cosmochimica Acta, 46, 2363,
  \dodoi{10.1016/0016-7037(82)90208-3}

\bibitem[{Anders \& Grevesse(1989)}]{anders_abundances_1989}
Anders, E., \& Grevesse, N. 1989, Geochimica et Cosmochimica Acta, 53, 197,
  \dodoi{10.1016/0016-7037(89)90286-X}

\bibitem[{Arnaud(1996)}]{arnaud_xspec:_1996}
Arnaud, K.~A. 1996, in Astronomical {Society} of the {Pacific} {Conference}
  {Series}, Vol. 101, 17.
\newblock \url{http://adsabs.harvard.edu/abs/1996ASPC..101...17A}

\bibitem[{Balucinska-Church \&
  McCammon(1992)}]{balucinska-church_photoelectric_1992}
Balucinska-Church, M., \& McCammon, D. 1992, The Astrophysical Journal, 400,
  699, \dodoi{10.1086/172032}

\bibitem[{Bellm \& Vaillancourt(2005)}]{bellm_origins_2005}
Bellm, E.~C., \& Vaillancourt, J.~E. 2005, The Astrophysical Journal, 622, 959,
  \dodoi{10.1086/428385}

\bibitem[{Bienayme {et~al.}(1987)Bienayme, Robin, \&
  Creze}]{bienayme_mass_1987}
Bienayme, O., Robin, A.~C., \& Creze, M. 1987, Astronomy and Astrophysics, 180,
  94

\bibitem[{Burrows \& Mendenhall(1991)}]{burrows_soft_1991}
Burrows, D.~N., \& Mendenhall, J.~A. 1991, Nature, 351, 629,
  \dodoi{10.1038/351629a0}

\bibitem[{Burrows \& Mendenhall(1993)}]{burrows_probing_1993}
---. 1993, Advances in Space Research, 13, 83,
  \dodoi{10.1016/0273-1177(93)90100-P}

\bibitem[{Cravens(1997)}]{cravens_comet_1997}
Cravens, T.~E. 1997, Geophysical Research Letters, 24, 105,
  \dodoi{10.1029/96GL03780}

\bibitem[{Crowder {et~al.}(2012)Crowder, Barger, Brandl, Eckart, Galeazzi,
  Kelley, Kilbourne, McCammon, Pfendner, Porter, Rocks, Szymkowiak, \&
  Teplin}]{crowder_observed_2012}
Crowder, S.~G., Barger, K.~A., Brandl, D.~E., {et~al.} 2012, The Astrophysical
  Journal, 758, 143, \dodoi{10.1088/0004-637X/758/2/143}

\bibitem[{Das {et~al.}(2019)Das, Mathur, Nicastro, \&
  Krongold}]{das_discovery_2019}
Das, S., Mathur, S., Nicastro, F., \& Krongold, Y. 2019, arXiv:1907.07176
  [astro-ph]

\bibitem[{Defay {et~al.}(2013)Defay, Morgan, McCammon, Wulf, Andrianarijaona,
  Fogle, Seely, Draganić, \& Havener}]{defay_x-ray_2013}
Defay, X., Morgan, K., McCammon, D., {et~al.} 2013, Physical Review A, 88,
  052702, \dodoi{10.1103/PhysRevA.88.052702}

\bibitem[{Ferrière(2001)}]{ferriere_interstellar_2001}
Ferrière, K.~M. 2001, Reviews of Modern Physics, 73, 1031,
  \dodoi{10.1103/RevModPhys.73.1031}

\bibitem[{Fogle {et~al.}(2014)Fogle, Wulf, Morgan, McCammon, Seely, Draganić,
  \& Havener}]{fogle_x-ray-emission_2014}
Fogle, M., Wulf, D., Morgan, K., {et~al.} 2014, Physical Review A, 89, 042705,
  \dodoi{10.1103/PhysRevA.89.042705}

\bibitem[{Galeazzi {et~al.}(2014)Galeazzi, Chiao, Collier, Cravens, Koutroumpa,
  Kuntz, Lallement, Lepri, McCammon, Morgan, Porter, Robertson, Snowden,
  Thomas, Uprety, Ursino, \& Walsh}]{galeazzi_origin_2014}
Galeazzi, M., Chiao, M., Collier, M.~R., {et~al.} 2014, Nature, 512, 171,
  \dodoi{10.1038/nature13525}

\bibitem[{Giampapa {et~al.}(1996)Giampapa, Rosner, Kashyap, Fleming, Schmitt,
  \& Bookbinder}]{giampapa_coronae_1996}
Giampapa, M.~S., Rosner, R., Kashyap, V., {et~al.} 1996, The Astrophysical
  Journal, 463, 707, \dodoi{10.1086/177284}

\bibitem[{Greenwood {et~al.}(2001)Greenwood, Williams, Smith, \&
  Chutjian}]{greenwood_experimental_2001}
Greenwood, J.~B., Williams, I.~D., Smith, S.~J., \& Chutjian, A. 2001, Physical
  Review A, 63, 062707, \dodoi{10.1103/PhysRevA.63.062707}

\bibitem[{Gupta \& Galeazzi(2009)}]{gupta_contribution_2009}
Gupta, A., \& Galeazzi, M. 2009, The Astrophysical Journal, 702, 270,
  \dodoi{10.1088/0004-637X/702/1/270}

\bibitem[{Henley \& Shelton(2015)}]{henley_xmm-newtonandsuzakux-ray_2015}
Henley, D.~B., \& Shelton, R.~L. 2015, The Astrophysical Journal, 808, 22,
  \dodoi{10.1088/0004-637X/808/1/22}

\bibitem[{Koutroumpa(2007)}]{koutroumpa_etude_2007}
Koutroumpa, D. 2007, PhD thesis, Pierre and Marie Curie University, Paris,
  France

\bibitem[{Koutroumpa {et~al.}(2009)Koutroumpa, Lallement, Raymond, \&
  Kharchenko}]{koutroumpa_solar_2009}
Koutroumpa, D., Lallement, R., Raymond, J.~C., \& Kharchenko, V. 2009, The
  Astrophysical Journal, 696, 1517, \dodoi{10.1088/0004-637X/696/2/1517}

\bibitem[{Kuntz \& Snowden(2000)}]{kuntz_deconstructing_2000}
Kuntz, K.~D., \& Snowden, S.~L. 2000, The Astrophysical Journal, 543, 195,
  \dodoi{10.1086/317071}

\bibitem[{Kuntz \& Snowden(2001)}]{kuntz_contribution_2001}
---. 2001, The Astrophysical Journal, 554, 684, \dodoi{10.1086/321421}

\bibitem[{Lei {et~al.}(2009)Lei, Shelton, \& Henley}]{lei_determining_2009}
Lei, S., Shelton, R.~L., \& Henley, D.~B. 2009, The Astrophysical Journal, 699,
  1891, \dodoi{10.1088/0004-637X/699/2/1891}

\bibitem[{Liu {et~al.}(2016)Liu, Galeazzi, \& Ursino}]{liu_evidence_2016}
Liu, W., Galeazzi, M., \& Ursino, E. 2016, The Astrophysical Journal, 816, 82,
  \dodoi{10.3847/0004-637X/816/2/82}

\bibitem[{Liu {et~al.}(2017)Liu, Chiao, Collier, Cravens, Galeazzi, Koutroumpa,
  Kuntz, Lallement, Lepri, McCammon, Morgan, Porter, Snowden, Thomas, Uprety,
  Ursino, \& Walsh}]{liu_structure_2017}
Liu, W., Chiao, M., Collier, M.~R., {et~al.} 2017, The Astrophysical Journal,
  834, 33, \dodoi{10.3847/1538-4357/834/1/33}

\bibitem[{Masui {et~al.}(2009)Masui, Mitsuda, Yamasaki, Takei, Kimura, Yoshino,
  \& McCammon}]{masui_nature_2009}
Masui, K., Mitsuda, K., Yamasaki, N.~Y., {et~al.} 2009, Publications of the
  Astronomical Society of Japan, 61, S115, \dodoi{10.1093/pasj/61.sp1.S115}

\bibitem[{McCammon {et~al.}(2002)McCammon, Almy, Apodaca, Bergmann~Tiest, Cui,
  Deiker, Galeazzi, Juda, Lesser, Mihara, Morgenthaler, Sanders, Zhang,
  Figueroa-Feliciano, Kelley, Moseley, Mushotzky, Porter, Stahle, \&
  Szymkowiak}]{mccammon_high_2002}
McCammon, D., Almy, R., Apodaca, E., {et~al.} 2002, The Astrophysical Journal,
  576, 188, \dodoi{10.1086/341727}

\bibitem[{McCammon {et~al.}(2008)McCammon, Barger, Brandl, Brekosky, Crowder,
  Gygax, Kelley, Kilbourne, Lindeman, Porter, Rocks, \&
  Szymkowiak}]{mccammon_x-ray_2008}
McCammon, D., Barger, K., Brandl, D.~E., {et~al.} 2008, Journal of Low
  Temperature Physics, 151, 715, \dodoi{10.1007/s10909-008-9734-5}

\bibitem[{Pearson(1900)}]{f.r.s_x._1900}
Pearson, K. F. R.~S. 1900, The London, Edinburgh, and Dublin Philosophical
  Magazine and Journal of Science, 50, 157, \dodoi{10.1080/14786440009463897}

\bibitem[{Ponti {et~al.}(2019)Ponti, Hofmann, Churazov, Morris, Haberl, Nandra,
  Terrier, Clavel, \& Goldwurm}]{ponti_x-ray_2019}
Ponti, G., Hofmann, F., Churazov, E., {et~al.} 2019, Nature, 567, 347,
  \dodoi{10.1038/s41586-019-1009-6}

\bibitem[{Robertson \& Cravens(2003)}]{robertson_spatial_2003}
Robertson, I.~P., \& Cravens, T.~E. 2003, Journal of Geophysical Research:
  Space Physics, 108, \dodoi{10.1029/2003JA009873}

\bibitem[{Rosner {et~al.}(1981)Rosner, Avni, Bookbinder, Giacconi, Golub,
  Harnden, Maxson, Topka, \& Vaiana}]{rosner_stellar_1981}
Rosner, R., Avni, Y., Bookbinder, J., {et~al.} 1981, The Astrophysical Journal,
  249, L5, \dodoi{10.1086/183646}

\bibitem[{Sanders {et~al.}(2001)Sanders, Edgar, Kraushaar, McCammon, \&
  Morgenthaler}]{sanders_spectra_2001}
Sanders, W.~T., Edgar, R.~J., Kraushaar, W.~L., McCammon, D., \& Morgenthaler,
  J.~P. 2001, The Astrophysical Journal, 554, 694, \dodoi{10.1086/321424}

\bibitem[{Savage \& Sembach(1996)}]{savage_interslar_1996}
Savage, B.~D., \& Sembach, K.~R. 1996, Annual Review of Astronomy and
  Astrophysics, 34, 279, \dodoi{10.1146/annurev.astro.34.1.279}

\bibitem[{Schlegel {et~al.}(1998)Schlegel, Finkbeiner, \&
  Davis}]{schlegel_maps_1998}
Schlegel, D.~J., Finkbeiner, D.~P., \& Davis, M. 1998, The Astrophysical
  Journal, 500, 525, \dodoi{10.1086/305772}

\bibitem[{Scholz \& Stephens(1987)}]{scholz_k-sample_1987}
Scholz, F.~W., \& Stephens, M.~A. 1987, Journal of the American Statistical
  Association, 82, 918, \dodoi{10.2307/2288805}

\bibitem[{Seely {et~al.}(2017)Seely, Andrianarijaona, Wulf, Morgan, McCammon,
  Fogle, Stancil, Zhang, \& Havener}]{seely_line_2017}
Seely, D.~G., Andrianarijaona, V.~M., Wulf, D., {et~al.} 2017, Physical Review
  A, 95, 052704, \dodoi{10.1103/PhysRevA.95.052704}

\bibitem[{Smith {et~al.}(2001)Smith, Brickhouse, Liedahl, \&
  Raymond}]{smith_collisional_2001}
Smith, R.~K., Brickhouse, N.~S., Liedahl, D.~A., \& Raymond, J.~C. 2001, The
  Astrophysical Journal Letters, 556, L91, \dodoi{10.1086/322992}

\bibitem[{Smith {et~al.}(2012)Smith, Foster, \&
  Brickhouse}]{smith_approximating_2012}
Smith, R.~K., Foster, A.~R., \& Brickhouse, N.~S. 2012, Astronomische
  Nachrichten, 333, 301, \dodoi{10.1002/asna.201211673}

\bibitem[{Smith {et~al.}(2014)Smith, Foster, Edgar, \&
  Brickhouse}]{smith_resolving_2014}
Smith, R.~K., Foster, A.~R., Edgar, R.~J., \& Brickhouse, N.~S. 2014, The
  Astrophysical Journal, 787, 77, \dodoi{10.1088/0004-637X/787/1/77}

\bibitem[{Smith {et~al.}(2007)Smith, Bautz, Edgar, Fujimoto, Hamaguchi, Hughes,
  Ishida, Kelley, Kilbourne, Kuntz, McCammon, Miller, Mitsuda, Mukai,
  Plucinsky, Porter, Snowden, Takei, Terada, Tsuboi, \&
  Yamasaki}]{smith_suzaku_2007}
Smith, R.~K., Bautz, M.~W., Edgar, R.~J., {et~al.} 2007, Publications of the
  Astronomical Society of Japan, 59, S141, \dodoi{10.1093/pasj/59.sp1.S141}

\bibitem[{Snowden {et~al.}(1998)Snowden, Egger, Finkbeiner, Freyberg, \&
  Plucinsky}]{snowden_progress_1998}
Snowden, S.~L., Egger, R., Finkbeiner, D.~P., Freyberg, M.~J., \& Plucinsky,
  P.~P. 1998, The Astrophysical Journal, 493, 715, \dodoi{10.1086/305135}

\bibitem[{Snowden {et~al.}(1997)Snowden, Egger, Freyberg, McCammon, Plucinsky,
  Sanders, Schmitt, Trümper, \& Voges}]{snowden_rosat_1997}
Snowden, S.~L., Egger, R., Freyberg, M.~J., {et~al.} 1997, The Astrophysical
  Journal, 485, 125, \dodoi{10.1086/304399}

\bibitem[{Uprety {et~al.}(2016)Uprety, Chiao, Collier, Cravens, Galeazzi,
  Koutroumpa, Kuntz, Lallement, Lepri, Liu, McCammon, Morgan, Porter, Prasai,
  Snowden, Thomas, Ursino, \& Walsh}]{uprety_solar_2016}
Uprety, Y., Chiao, M., Collier, M.~R., {et~al.} 2016, The Astrophysical
  Journal, 829, 83, \dodoi{10.3847/0004-637X/829/2/83}

\bibitem[{von Steiger {et~al.}(2000)von Steiger, Schwadron, Fisk, Geiss,
  Gloeckler, Hefti, Wilken, Wimmer-Schweingruber, \&
  Zurbuchen}]{von_steiger_composition_2000}
von Steiger, R., Schwadron, N.~A., Fisk, L.~A., {et~al.} 2000, Journal of
  Geophysical Research: Space Physics, 105, 27217, \dodoi{10.1029/1999JA000358}

\bibitem[{Weisskopf {et~al.}(2010)Weisskopf, Guainazzi, Jahoda, Shaposhnikov,
  O’Dell, Zavlin, {C. Wilson-Hodge}, \& Elsner}]{weisskopf_calibrations_2010}
Weisskopf, M.~C., Guainazzi, M., Jahoda, K., {et~al.} 2010, The Astrophysical
  Journal, 713, 912, \dodoi{10.1088/0004-637X/713/2/912}

\bibitem[{Wulf {et~al.}(2016)Wulf, Jaeckel, McCammon, \&
  Morgan}]{wulf_technique_2016}
Wulf, D., Jaeckel, F., McCammon, D., \& Morgan, K.~M. 2016, Journal of Low
  Temperature Physics, 184, 431, \dodoi{10.1007/s10909-015-1445-0}

\bibitem[{Wulf {et~al.}(2019)Wulf, Jaeckel, McCammon, \&
  Morgan}]{wulf_technique_2019}
---. 2019, Submitted to XXXX

\bibitem[{Yao \& Wang(2007)}]{yao_nonisothermality_2007}
Yao, Y., \& Wang, Q.~D. 2007, The Astrophysical Journal, 658, 1088,
  \dodoi{10.1086/512003}

\bibitem[{Yao {et~al.}(2009)Yao, Wang, Hagihara, Mitsuda, McCammon, \&
  Yamasaki}]{yao_x-ray_2009}
Yao, Y., Wang, Q.~D., Hagihara, T., {et~al.} 2009, The Astrophysical Journal,
  690, 143, \dodoi{10.1088/0004-637X/690/1/143}

\bibitem[{Yoshino {et~al.}(2009)Yoshino, Mitsuda, Yamasaki, Takei, Hagihara,
  Masui, Bauer, McCammon, Fujimoto, Daniel~Wang, \& Yao}]{yoshino_energy_2009}
Yoshino, T., Mitsuda, K., Yamasaki, N.~Y., {et~al.} 2009, Publications of the
  Astronomical Society of Japan, 61, 805, \dodoi{10.1093/pasj/61.4.805}

\end{thebibliography}

\end{document}